\documentclass[sigconf,review=false,nonacm]{acmart} 
\usepackage{listings}
\usepackage{todonotes}
\usepackage{subcaption}
\usepackage{algorithm}
\usepackage[noend]{algpseudocode}
\usepackage{xltabular} 
\newcommand{\var}[1]{\ensuremath{\mathit{#1}}}
\newtheorem{property}{Property}

\floatname{algorithm}{State Machine}
\newcommand{\component}{\textbf{State}}
\newcommand{\componentcolon}{\textbf{State:}}
\newcommand{\Component}{\item[\componentcolon]}

\newcommand{\initial}{\textbf{Initialize}}
\newcommand{\Initial}[1]{\item[\initial(#1):]}

\newcommand{\Move}{\item[\textbf{Move}:]}

\newcommand{\kYes}{\textsc{yes}}
\newcommand{\kNo}{\textsc{no}}

\newcommand{\longAccount}{\ensuremath{\mathit{LongAccount}}}
\newcommand{\shortAccount}{\ensuremath{\mathit{ShortAccount}}}
\newcommand{\AliceYes}{\ensuremath{\mathit{AliceYes}}}
\newcommand{\AllDone}{\ensuremath{\mathit{AllDone}}}
\newcommand{\BobYes}{\ensuremath{\mathit{BobYes}}}

\newcommand{\Self}{\ensuremath{\mathit{Self}}}
\newcommand{\Sender}{\ensuremath{\mathit{Sender}}}

\newcommand{\account}{\ensuremath{\mathit{account}}}
\newcommand{\addr}{\ensuremath{\mathit{addr}}}
\newcommand{\age}{\ensuremath{\mathit{age}}}

\newcommand{\bid}{\ensuremath{\mathit{bid}}}
\newcommand{\buffer}{\ensuremath{\mathit{buffer}}}

\newcommand{\defundVote}{\ensuremath{\mathit{defundVote}}}
\newcommand{\defund}{\ensuremath{\mathit{defund}}}
\newcommand{\delivered}{\ensuremath{\mathit{delivered}}}
\newcommand{\deliver}{\ensuremath{\mathit{deliver}}}
\newcommand{\enabled}{\ensuremath{\mathit{enabled}}}

\newcommand{\funded}{\ensuremath{\mathit{funded}}}
\newcommand{\fund}{\ensuremath{\mathit{fund}}}

\newcommand{\halt}{\ensuremath{\mathbf{halt}}}

\newcommand{\leader}{\ensuremath{\mathit{leader}}}
\newcommand{\live}{\ensuremath{\mathit{live}}}
\newcommand{\machine}{\ensuremath{\mathit{machine}}}
\newcommand{\moves}{\ensuremath{\mathit{moves}}}

\newcommand{\noVotes}{\ensuremath{\mathit{noVotes}}}
\newcommand{\now}{\ensuremath{\mathit{now}}}

\newcommand{\ready}{\ensuremath{\mathit{ready}}}
\newcommand{\redeem}{\ensuremath{\mathit{redeem}}}

\newcommand{\sealed}{\ensuremath{\mathit{sealed}}}

\newcommand{\seen}{\ensuremath{\mathit{seen}}}
\newcommand{\send}{\ensuremath{\mathit{send}}}
\newcommand{\sig}{\ensuremath{\mathit{sig}}}
\newcommand{\skipMove}{\ensuremath{\mathit{Skip}}}
\newcommand{\startTime}{\ensuremath{\mathit{startTime}}}

\newcommand{\topUpVerified}{\ensuremath{\mathit{topUpVerified}}}
\newcommand{\topUp}{\ensuremath{\mathit{topUp}}}

\newcommand{\util}{\ensuremath{\mathit{util}}}
\newcommand{\verifyAccounts}{\ensuremath{\mathit{verifyAccounts}}}

\newcommand{\voted}{\ensuremath{\mathit{voted}}}

\newcommand{\yesVotes}{\ensuremath{\mathit{yesVotes}}}

\renewcommand{\succ}{\ensuremath{\mathit{succ}}}

\newcommand{\Alice}{\ensuremath{\text{Alice}}}
\newcommand{\Bob}{\ensuremath{\text{Bob}}}

\newcommand{\ducat}{\ensuremath{\text{ducat}}}
\newcommand{\false}{\ensuremath{\mathit{false}}}
\newcommand{\florin}{\ensuremath{\text{florin}}}

\newcommand{\token}{\ensuremath{\text{token}}}
\newcommand{\true}{\ensuremath{\mathit{true}}}

\def\cADDR{\ensuremath{{\mathcal{ADDR}}}}
\def\cASSET{\ensuremath{{\mathcal{ASSET}}}}

\def\cREQ{\ensuremath{{\mathcal{REQ}}}}
\def\cPS{\ensuremath{{\mathcal{PS}}}}

\def\cA{\ensuremath{{\mathcal A}}}         
   \def\cF{\ensuremath{{\mathcal F}}}      
         
\def\cM{\ensuremath{{\mathcal M}}}         
      \def\cS{\ensuremath{{\mathcal S}}}

\newcommand{\set}[1]{\left\{ #1 \right\}}

\newcommand{\eqnlabel}[1]{\label{eq:#1}}

\newcommand{\thmlabel}[1]{\label{thm:#1}}

\newcommand{\seclabel}[1]{\label{sec:#1}}
\newcommand{\nakedsecref}[1]{\ref{sec:#1}}
\newcommand{\secref}[1]{Section~\nakedsecref{#1}}

\newcommand{\linlabel}[1]{\label{line:#1}}
\newcommand{\nakedlineref}[1]{\ref{line:#1}}
\newcommand{\linref}[1]{Line~\nakedlineref{#1}}
\newcommand{\linrefrange}[2]{Lines~\nakedlineref{#1}-\nakedlineref{#2}}

\newcommand{\gamelabel}[1]{\label{game:#1}}
\newcommand{\nakedgameref}[1]{\ref{game:#1}}
\newcommand{\gameref}[1]{Algorithm~\nakedgameref{#1}}

\newcommand{\algolabel}[1]{\label{algo:#1}}
\newcommand{\nakedalgoref}[1]{\ref{algo:#1}}
\newcommand{\algoref}[1]{Algorithm~\nakedalgoref{#1}}

\newcommand{\bbR}{\ensuremath{\mathbb{R}}}

\graphicspath{{./pictures/}}

\title{Cross-Chain State Machine Replication}

\author{Yingjie Xue}
\affiliation{
   \institution{Computer Science Dept., Brown University}
   \city{Providence}
   \state{RI}
   \country{USA}}
\author{Maurice Herlihy}
       \affiliation{
   \institution{Computer Science Dept., Brown University}
   \city{Providence}
   \state{RI}
   \country{USA}}

\begin{abstract}
This paper considers the classical state machine replication (SMR) problem
in a distributed system model inspired by cross-chain exchanges.
We propose a novel SMR protocol adapted for this model.
Each state machine transition takes $O(n)$ message delays,
where $n$ is the number of active participants,
of which \emph{any number} may be Byzantine.
This protocol makes novel use of path signatures \cite{Herlihy2018}
to keep replicas consistent.
This protocol design cleanly separates application logic from fault-tolerance,
providing a systematic way to replace complex \emph{ad-hoc} cross-chain protocols
with a more principled approach.

\end{abstract}

\begin{document}
\maketitle

\section{Introduction}
\seclabel{intro}
In the \emph{state machine replication} (SMR) problem,
a service, modeled as a state machine,
is replicated across multiple servers to provide fault tolerance.
SMR has been studied in models of computation
subject to crash failures \cite{Lamport1998,OngaroO2014} and Byzantine failures \cite{PBFT},
in both synchronous \cite{synchot} and asynchronous \cite{duan2018beat} timing models.

This paper proposes an SMR protocol for a model of computation
inspired by, but not limited to, transactions that span multiple blockchains.
The service's state is replicated across multiple automata.
These replicas model smart contracts on blockchains: they are \emph{trustworthy},
responding correctly to requests,
but \emph{passive}, 
meaning they undergo state changes only in response to outside requests.
Like smart contracts,
replicas cannot communicate directly with other replicas or observe their states.
Active \emph{agents} initiate replica state changes
by communicating with the replicas over authenticated channels.
Agents model blockchain users: any number of them may be \emph{Byzantine},
eager to cheat other agents in arbitrary (but computationally-bounded) ways.

This SMR protocol guarantees \emph{safety},
meaning that Byzantine agents cannot victimize honest agents,
and \emph{liveness},
meaning that if all agents are honest,
then all replicas change state correctly.

Although cross-chain SMR and conventional SMR have (essentially) the same formal structure,
their motivations differ in important ways.
Conventional SMR embraces distribution to make services fault-tolerant.
By contrast, individual blockchains are already fault-tolerant.
Instead, cross-chain SMR is motivated by the need for
\emph{interoperability} across multiple independent chains.
For example,
suppose Alice and Bob have euro accounts on a chain run by the European Central Bank,
and dollar accounts on a chain run by the Federal Reserve.
They agree to a trade:
Alice will transfer some euros to Bob if Bob transfers some dollars to Alice.
Realistically, however,
Alice and Bob will never be able execute their trade on a single chain
because the dollar chain and the euro chain will always be distinct for political reasons.
They could use an \emph{ad hoc} cross-chain swap protocol~\cite{Herlihy2018,tiersnolan},
but an SMR protocol has a cleaner structure,
and generalizes more readily to more complex exchanges.
So Alice and Bob codify their trade as a simple, centralized state machine that
credits and debits their accounts.
They place a state machine replica on each chain,
and execute the trade through an SMR protocol that keeps those replicas consistent.
While both replicas formally execute the same steps,
the euro chain replica actually transfers the euros,
the dollar chain replica actually transfers the dollars,
and the SMR protocol ensures these transfers happen atomically.

A conceptual benefit of SMR over \emph{ad-hoc} protocols is separation of concerns.
Expressing a complex financial exchange as a (non-distributed) state machine frees the protocol designer to focus on the exchange's incentives, payoffs, and equilibria,
without simultaneously having to reason about timeout duration, missing or corrupted communication.

Prior SMR protocols assume some fraction of the participants
(usually more than one-half or two-thirds) to be non-faulty.
By contrast, for cross-chain applications it does not make sense to
assume a limit on the number of Byzantine agents.
Instead, this model's SMR protocol protects agents
who honestly follow the protocol from those who don't,
all while ensuring progress when enough agents are honest.

This paper makes the following contributions.
We are the first to consider the classical SMR coordination problem
in a distributed system model inspired by cross-chain exchanges.
The model itself is a formalization of models implicit in earlier,
more applied works~\cite{bitcoin,GarayKL2015,Herlihy2018,HerlihyLS2021}.
Fundamental coordination problems in this model have received little formal analysis.
We propose a novel SMR protocol adapted for this model.
Each state machine transition takes $O(n)$ message delays,
where $n$ is the number of agents,
of which \emph{any number} may be Byzantine.
This protocol makes novel use of path signatures \cite{Herlihy2018}
to keep replicas consistent.
This SMR structure cleanly separates application logic from fault-tolerance,
providing a systematic way to replace complex \emph{ad-hoc} cross-chain protocols
with a more principled approach.

This paper is organized as follows.
\secref{model} describes the cross-chain model of computation,
\secref{games} gives examples of automata representing various kinds of cross-chain exchanges,
\secref{smr} describes our cross-chain SMR protocol,
\secref{remarks} discusses optimizations and extensions,
and \secref{related} surveys related work.

\section{Model of Computation}
\seclabel{model}
Our model is motivated by today's blockchains and smart contracts, 
but it does not assume any specific blockchain technology,
or even blockchains as such.
Instead, we focus on computational abstractions central to any
systematic approach to exchanges of value among untrusting agents,
whatever technology underlies the shared ledger.

The system consists of a set of communicating automata.
An automaton is either an active, untrusted \emph{agent},
or a passive, trusted \emph{replica}.
An agent automaton models a blockchain client such as a person or an organization.
Agents are untrusted because they model untrusted blockchain clients.
A replica automaton models a \emph{smart contract} (or \emph{contract}),
a chain-resident program that manipulates ledger state.
Contract code and state are public,
and that code is reliably executed by validators who reach consensus on each call.
Replicas are trusted because they model trusted contracts.

Reflecting the limitations of today's blockchains,
agents communicate only with replicas
(clients can only call contract functions),
and replicas do not communicate with other replicas
(contracts on distinct chains cannot communicate).
Replica $A$ can learn of a state change at replica $B$
only if some agent explicitly informs $A$ of $B$'s new state.
Of course, $A$ must decide whether that agent is telling the truth. 

Like prior work~\cite{Herlihy2018,HerlihyLS2021,tiersnolan,ZakharyAE2019},
we assume a \emph{synchronous} network model where communication time is known and bounded.
There is a time bound $\Delta > 0$,
such that when an agent initiates a state change at a replica,
that change will observed by all agents within time $\Delta$.
In our pseudocode examples, the function $\var{now()}$ returns the current time.
We do not assume clocks are perfectly synchronized,
only that clock drifts are kept small in comparison to $\Delta$.

We make standard cryptographic assumptions.
Each agent has a public and a private key,
with public keys known to all.
Messages are signed so they cannot be forged,
and they include single-use labels (``nonces'')
so they cannot be replayed.

The agents participating in an exchange agree on a common \emph{protocol}:
rules that dictate when to request replica state changes.
Instead of distinguishing between faulty and non-faulty agents,
as in classical SMR models,
we distinguish only between \emph{compliant} (i.e. \emph{honest}) agents
who honestly follow the common protocol,
and \emph{deviating} (i.e. \emph{Byzantine}) agents who do not.
Unlike prior SMR models,
which require some fraction of the agents to be compliant,
we tolerate any number of Byzantine agents\footnote{
If all agents are Byzantine, then correctness becomes vacuous.}.

\section{State Machines}
\seclabel{games}
\floatname{algorithm}{State Machine}
Because applications such as cross-chain auctions or swaps
are typically structured as multi-step protocols where
agents take turns transferring assets in and out of escrow accounts
\cite{HerlihyLS2021,tiersnolan,ZakharyAE2019},
the state machine is structured as a multi-agent game.
For simplicity, agents make moves in round-robin order.
(In practice, agents can sometimes skip moves or move concurrently.)

\sloppy Formally, a \emph{game} is defined by a decision tree
$G=(\cA,\cM,\cS,\cF, \moves, \enabled, \succ, \util)$, where
$\cA$ is a set of $n > 1$ \emph{agents},
$\cM$ is a set of \emph{moves},
$\cS$ is a set of \emph{non-final states},
and $\cF$ is a set of \emph{final states} disjoint from $\cS$.
$\cS$ includes a distinguished \emph{initial state} $s_0$.
The function
$\moves: \cS \to 2^\cM$
defines which moves are enabled at each non-final state,
$\enabled: \cS \to \cA$
defines which agent chooses the next move at each non-final state,
$\succ: \cS \times \cM \to \cS \cup \cF$,
defines which state is reached following a move in a non-final state.
This successor function induces a tree structure on states:
for $s_1,s_2 \in \cS$ and $m_1,m_2\in \cM$,  
if $\succ(s_1,m_1)=\succ(s_2,m_2)$ then $s_1=s_2$ and $m_1=m_2$.
Finally,
the utility function $\util: \cA \times \cF \to \bbR^n$
is given by a vector of real-valued functions on final states, indexed by agent:
$(\util_P: \cF \to \bbR \;|\; P \in \cA)$.
For each agent $P \in \cA$ and state $z \in \cF$,
$\util_P(z)$ measures $P$'s preference for $z$ compared to its preference for
the initial state.
Informally,
$\util_P(z)$ is negative for states where $P$ ends up ``worse off'' than it started,
positive for states where $P$ ends up ``better off'',
and zero for states where $P$ is indifferent.
An \emph{execution} from state $s_0$ is a sequence
$(s_0, P_1, \mu_1, \ldots, s_{i-1}, P_i, \mu_i, \ldots, s_k)$
where each $P_i = \enabled(s_{i-1})$, $\mu_i \in \moves(s_{i-1})$,
$s_i=\succ(s_{i-1},\mu_i)$, and $s_k \in \cF$.
We divide executions into \emph{rounds}:
move $\mu_i$ takes place at round $i$.
Game trees are finite and deterministic, hence so are executions.

Not all game trees make sense as abstract state machines.
We are not interested in games like chess or poker where
one agent's gain is another agent's loss.
Instead, we are interested in games where all agents stand to gain.
A \emph{protocol} $\Pi: \cS \to 2^{\cM}$ is a rule for choosing among enabled moves.
As mentioned,
agents that follow the protocol are \emph{compliant},
while those who do not are \emph{deviating}.
More precisely,
$P$ is compliant in an execution
$(s_0, P_1, \mu_1, \ldots, s_{i-1}, P_i, \mu_i, \ldots, s_k)$
if it follows the protocol:
if $P = \enabled(s_{i-1})$, then $\mu_i \in  \Pi(s_{i-1})$.
An execution is compliant if every agent follows the protocol:
for $i \in 1 \ldots k$, $\mu_i \in \Pi(s_{i-1})$.

A \emph{mutually-beneficial protocol} guarantees:
\begin{itemize}
\item
  \emph{Liveness}:
  Every compliant execution
  leads to a final state $z$ where $\util_P(z) > 0$ for all $P \in \cA$.
\item
  \emph{Safety}:
  Every execution in which agent $P$ is compliant
  leads to a final state $z$ where $\util_P(z) \geq 0$.
\end{itemize}
The first condition says that if all agents are compliant,
they all end up strictly better off.
The second says that a compliant agent will never end up worse off,
even if others deviate.
Establishing these properties is the responsibility of the game designer,
and preserving them is the responsibility
of the SMR protocol.

Both agents and the state machine itself can own and exchange \emph{assets}.
We keep track of ownership using \emph{addresses}:
each agent $P$ has an \emph{address}, $\addr(P)$,
and the state machine has an address $\Self$.
Let $\cADDR$ be the domain of addresses,
and $\cASSET$ the domain of assets.

We represent state machines in procedural pseudocode.
The block marked \component{} defines the machine's state components.
The state includes an \emph{account} map,
$\account: \cADDR  \times \cASSET \to \mathbb{Z}$,
mapping addresses and assets to account balances.
We will often abuse notation by writing
$\account(P,A)$ in place of $\account(\addr(P),A)$
when there is no danger of confusion.
Other state components may include counters, flags,
or other bookkeeping structures.

At the start of the state machine execution,
the agents \emph{initialize} the state executing
the block marked \initial$(\ldots)$.
An agent triggers a state transition by issuing a \emph{move},
which may take arguments.
Each move has an implicit $\Sender$ argument that keeps track
of which agent originated that move.
In examples, a move is defined by a \textbf{Move} block,
which checks preconditions and enforces postconditions.
To capture the Byzantine nature of agents,
every non-final state has an enabled \skipMove{} move,
which leaves the state unchanged,
except for moving on to the next turn.
(Usually, \skipMove{} deviates from the protocol.)
The keyword \halt{} ends the execution for the sender.

The example state machines illustrated in this section 
favor readability over precision when meanings are clear.
For clarity and brevity, 
we omit some routine sanity checks and error cases.
Our examples are all applications that exchange assets because
these are the applications that make the most sense for the cross-chain model.

\subsection{Example: Simple Swap}
\begin{algorithm}[htb]
\caption{Simple Swap}
\gamelabel{swap}
\begin{algorithmic}[1]
  \Component
  \State $\account: \cADDR \times \cASSET \to \mathbb{Z}$
  \State \AliceYes, \BobYes, \AllDone: bool := \false, \false, \false
  \Move Agree()\Comment{Each agent agrees to swap}
    \If{$\Sender = \Alice \wedge \account(\Alice,\florin) \geq 1$}\linlabel{swap:alice}
      \State $\AliceYes := \true$
    \ElsIf{$\Sender = \Bob \wedge \account(\Bob,\ducat) \geq 1$}\linlabel{swap:bob}
      \State $\BobYes := \true$
      \EndIf
  \Move Complete()\Comment{Any agent can complete the swap}\linlabel{swap:complete}
    \If{$\neg\AllDone$}\Comment{Not yet completed?}
      \If{$\AliceYes \wedge \BobYes$}\Comment{Both agents agreed?}
        \State $\account(\Alice,\florin) := \account(\Alice,\florin) - 1$
        \State $\account(\Bob,\florin) := \account(\Bob,\florin) + 1$
        \State $\account(\Bob,\ducat) := \account(\Bob,\ducat) - 1$
        \State $\account(\Alice,\ducat) := \account(\Alice,\ducat) + 1$
        \EndIf
      \State $\AllDone = true$
      \EndIf
    \State \halt
\end{algorithmic}
\end{algorithm}

\gameref{swap} shows pseudocode for a \emph{simple swap} state machine,
where Alice and Bob swap one of her florins for one of his ducats.
The block marked \textbf{State} defines the state components:
the accounts map and various control flags.
Each agent agrees to the swap (\linref{swap:alice}, \linref{swap:bob}),
checking that the caller has sufficient funds.
After both have agreed,
either agent can complete the transfers (\linref{swap:complete}).
If either agent tries to complete the transfer before both have agreed,
the transfer fails, and no assets are exchanged.

\subsection{Example: Decentralized Autonomous Organization (DAO)}
\begin{algorithm}[htb]
\caption{DAO State Machine}
\gamelabel{dao}
\begin{algorithmic}[1]
  \Component
  \State $\account: \cADDR \times \cASSET \to \mathbb{Z}$\Comment{Initially 0}
  \State $\yesVotes: \cA \to \mathbb{Z}, \noVotes: \cA \to \mathbb{Z}$\Comment{Initially 0}
  \State $\voted: \cA \to \set{\true,\false}$\Comment{Initially $\false$}
  \Initial{}
    \If{$\account(\Self,\florin) < 100$}\Comment{Make sure DAO has funds}\linlabel{dao:init}
      \State \halt
      \EndIf
  \Move VoteYes($k: \mathbb{Z}$)\Comment{LP casts $k$ votes}
    \If{$\Sender$ is enabled}\linlabel{dao:vote0}
      \If{$\account(\Sender,\token) \geq k$}\Comment{Sender has enough tokens}
        \State $\yesVotes(\Sender) := k$
        \State $\voted(\Sender) := \true$\linlabel{dao:vote1}
        \EndIf
      \EndIf
  \Move VoteNo($k: \mathbb{Z}$)
    $\ldots$\Comment{Symmetric with \var{VoteYes}}
  \Move $\skipMove()$
        \State do nothing\linlabel{dao:skip}
  \Move Resolve()
    \If{$\Sender$ is enabled and threshold voted \kYes}\Comment{Fund 100 to \Alice}\linlabel{dao:res0}
      \State $\account(\Self,\florin) := account(\Self,\florin) - 100$
      \State $\account(\Alice,\florin) := account(\Alice,\florin) + 100$
      \State \halt\linlabel{dao:res1}
      \EndIf
\end{algorithmic}
\end{algorithm}

Consider a venture fund organized as a \emph{decentralized autonomous organization} (DAO),
where liquidity providers (LPs) vote on how to invest their funds.
\gameref{dao} shows a state machine where
the DAO's LPs vote on whether to fund Alice's request for 100 florins.
Each LP holds some number of \emph{governance tokens},
each of which can be converted to a vote.
After the LPs vote, a \emph{director} tallies their votes,
and if there are enough \kYes{} votes, transfers the funds.
The state consists of accounts,
$\account: \cADDR \times \cASSET \to \mathbb{Z}$,
and maps $\yesVotes: \cA \to \mathbb{Z}$
and $\noVotes: \cA \to \mathbb{Z}$ counting \kYes{} and \kNo{} votes.

Initialization (\linref{dao:init}) ensures that
the DAO's own account is funded.
Each LP votes in turn whether to approve Alice's request
(\linrefrange{dao:vote0}{dao:vote1}).
(As discussed in \secref{remarks}, these votes could be concurrent.)
If an LP skips its turn, the tallies are unchanged (\linref{dao:skip}).
After every LP has had a chance to vote,
the director can ask for a resolution (\linrefrange{dao:res0}{dao:res1}).
If the caller is authorized and if a threshold number of votes were \kYes{}
(\linref{dao:res0}),
the funds are transferred to Alice from the DAO's account.
In either case, the execution ends.

\subsection{Example: Sealed-Bid Auction}
\begin{algorithm}[htb]
\caption{Sealed-Bid Auction State Machine}
\gamelabel{auction}
\begin{algorithmic}[1]
  \Component
  \State $\account: \cADDR \times \cASSET \to \mathbb{Z}$
  \State $\sealed: \cA \to \mathbb{Z}$ \Comment{Hash of bid + nonce}
  \State $\bid: \cA \to \mathbb{Z}$ \Comment{Unsealed bid}
  \Move SealedBid$(s)$\linlabel{auc:sealed}
    \If{$\Sender$ is enabled}\Comment{No moves out of turn}
      \State $\sealed(\Sender) := s$
      \EndIf
  \Move Unseal$(b: \mathbb{Z}, n: \mathbb{Z})$\linlabel{auc:unseal}
    \If{$\Sender$ is enabled}\Comment{No moves out of turn}
      \If{$H(b || n) = \sealed(\Sender)$}\Comment{Sealed bid checks out}
        \If{$\account(\Sender,\florin) \geq b$}\Comment{Bidder has enough money?}
          \State $\account(\Sender,\florin) := \account(\Sender,\florin) - b$
          \Comment{Transfer bid}
          \State $\account(\Self,\florin) := \account(\Self,\florin) + b$
          \State $\bid(\Sender) := b$
          \EndIf
        \EndIf
      \EndIf
  \Move Resolve()\Comment{Bidder collects NFT or refund}\linlabel{auc:resolve}
    \If{$\Sender$ is enabled and all bids unsealed}
  \If{ $(\forall P \in \cA)$
        $(
        \bid(\Sender) > \bid(P)
        \vee$ \\
          $(\bid(\Sender) = \bid(P) 
          \wedge \Sender > P)
        )$}\linlabel{auc:check}
        \State transfer NFT to $\Sender$\Comment{Sender won} 
      \Else \Comment{Sender lost, refund bid}        
        \State $\account(\Sender, \florin) := \account(\Sender, \florin) + \bid(\Sender)$
        \linlabel{auc:refund}
        \State $\account(\Self, \florin) := \account(\Self, \florin) - \bid(\Sender)$
        \EndIf
      \EndIf
\end{algorithmic}
\end{algorithm}

Consider a simple auction,
where bids are placed on one ledger for an NFT asset maintained on another.
As in the swap state machine,
each agent funds its account before the execution starts.
(In \secref{smr},
we explain how bidders might move funds
to the state machine while the execution is in progress.)

The state consists of the accounts,
a set of \emph{sealed bids},
$\sealed: \cA \to \mathbb{Z}$,
and a set of (plain-text) bids,
$\bid: \cA \to \mathbb{Z}$.
Each bidder conducts a simple commit-reveal protocol.
First,
the bidder issues a sealed bid
constructed by hashing its bid together with a nonce (\linref{auc:sealed}).
After all bidders have submitted their sealed bids,
each bidder submits their actual bid and nonce,
which is checked by the state machine (\linref{auc:unseal}).
After all bids have been received and unsealed,
any bidder can query the outcome by calling \emph{resolve()} (\linref{auc:resolve}).
If the sender's bid was highest (after breaking ties) (\linref{auc:check}),
the NFT is transferred,
and otherwise the sender's bid is refunded
from the replica's account(\linref{auc:refund}).

\section{State Machine Replication Protocol}
\seclabel{smr}
In this section,
we define an SMR protocol
by which multiple \emph{replica} automata
emulate a (centralized) state machine as defined in the previous section.

There are $n$ agents and $m$ assets,
where each asset is managed by its own replica.
Each replica maintains its own copy of the shared state.
The SMR protocol's job is to keep those copies consistent.
We assume that agents have some way to find one another,
to agree on the state machine defining their exchange,
and to initialize replicas that begin execution with synchronized clocks.

The core of the SMR protocol is a reliable delivery service
that ensures that the moves issued by the agents are delivered
to the replicas reliably, in order.
Reliable ordered delivery in the presence of Byzantine failures is
well-studied~\cite{Bracha1987,guerraouiKMPSV2020,MendesHT2012,SrikanthT1987},
but the cross-chain model requires new protocols
because the rules are different.
The principal difference is the asymmetry between \emph{agents},
active automata who cannot be trusted,
and \emph{replicas},
purely reactive automata who can observe only their own local states,
but who can be trusted to execute their own transitions correctly.

For example,
suppose the protocol calls for Alice to send a move
to replicas $A$ and $B$,
instructing them to transition to state $s$.
Each replica that receives the move validates Alice's signature,
checks that it is Alice's turn,
and that the move is enabled in the current state.

There are several ways Alice might deviate.
First, she might send her move to replica $A$ but not $B$.
In the SMR protocol, however, agents monitor one another.
Another compliant agent, Bob,
will notice that $B$ has not received Alice's move.
Bob will sign and relay that move from $A$ to $B$,
causing $B$ to receive that move at most $\Delta$ later than $A$.
As long as there is at least one compliant agent,
each move will be delivered to each replica within
a known duration.

Second, Alice might deviate by sending conflicting moves,
such as ``transition to $s$'' to $A$,
but ``transition to $s'$'' to $B$.
Here, too, Bob will notice the discrepancy
and relay both moves to $A$ and $B$,
presenting each replica with proof that Alice deviated.
Each replica will discard the conflicting moves,
acting as if Alice had skipped her turn.

Third, Alice might send a move to $A$ that is not
enabled in the current state.
Replica $A$ simply ignores that move,
acting as if Alice had skipped her turn.

Finally, Alice might not send her move to either replica.
Each replica that goes long enough without receiving a move
will act as if Alice had skipped her turn.
In short, reliable delivery has only two outcomes:
a valid move from Alice delivered to every replica,
or no valid move delivered, interpreted as a \skipMove,
all within a known duration.

To summarize,
the SMR protocol consists of three modules.
\begin{itemize}
\item 
  The \emph{front-end} automata (\algoref{front-end}), one for each agent,
  provide functions called by agents,
  including initial asset transfers into the state machine,
  the moves,
  and final asset transfers out of the state machine.
  (Every compliant agent is in charge of ensuring that final asset transfers take place.)

\item
  The \emph{relay} service (\algoref{relay}) guarantees that moves issued by front-ends
  are reliably delivered to the replicas
  as long as at least one agent is compliant.

\item
  The \emph{replicas} (\algoref{replica}), one for each asset, process function calls sent by agents from front-end,
  maintain copies of the state, and manage individual assets.
\end{itemize}

\subsection{Path Signatures}
A \emph{request} $req$ is a triple $(P, \mu, r)$,
used to indicate that agent $P$ requests move $\mu$ at the start of round $r$.
Let $\sig_P(P,m)$ denote the result of signing a message $m$ with $P$'s secret key.
A \emph{path} $p$ of length $k$ is a sequence $[P_1, \ldots, P_k]$ of distinct agents.
We use $[]$ for the empty sequence, and $[p,Q]$ to 
append $Q$ to the sequence $p$:
$[[P_1,\ldots,P_k], Q] = [P_1,\ldots,P_k, Q]$.
A \emph{path signature}~\cite{Herlihy2018,HerlihyLS2021} for $p$ is defined inductively:
\begin{equation*}
  \eqnlabel{path}
  p(P,\mu,r) :=
  \begin{cases}
    (P,\mu,r) &\text{If $p = []$,}\\
    \sig_Q(Q,q(P,\mu,r)) &\text{if $p = [q,Q]$}
  \end{cases}
\end{equation*}
\sloppy Informally, path signatures work as follows.
The $r^\text{th}$ round starts at time $t=(r-1)n\Delta$ after initialization.
Within time $t+\Delta$ after the start of round $r$,
a receiver accepts the path signature $[\Alice](\Alice, \mu, r)$ directly from \Alice.
Within time $t+2\Delta$,
a receiver accepts $[\Alice, \Bob](\Alice, \mu, r)$
originating from Alice and relayed through Bob,
and within time $t+k\Delta$,
a receiver accepts a message originating from Alice and relayed through $k-1$ distinct agents.
A path signature of length $k$ is \emph{live} for a duration of $k \Delta$ after $t$.
If no message is received for a duration of $n \Delta$ after $t$,
then no message was sent.

Define the following functions and predicates on path signatures of length $k$:
\begin{align*}
  \age(p(P, \mu, r)) &:= now()-(r-1)n\Delta \\
  \live(p(P, \mu, r)) &:= \age(p(P, \mu, r)) \leq k \Delta\\ 
  \ready(p(P, \mu, r)) &:= \age(p(P, \mu, r)) > n \Delta\\
\end{align*}
The $\age()$ function is the time elapsed since the start of the current round.
The SMR protocol uses $\live()$ to determine whether a message should be accepted by replicas,
and $\ready()$ to determine whether the accepted message's move can be applied.
A path signature is \emph{well-formed} if the signatures are valid
and the signers are distinct.
For brevity, replica pseudocode omits well-formedness checks.
We use $\cREQ=\cA \times \cM \times \mathbb{Z}$ for the domain of requests,
and $\cPS$ for the domain of path signatures.

\subsection{Reliable Delivery}
The $n$ agents act as \emph{senders} (indexed by $\cA$)
and the $m$ replicas act as \emph{receivers} (indexed by $\cASSET$).

Each receiver $A$ has a component: $\buffer_A: \cA \to 2^\cPS$,
where $\buffer_A(P)$ holds path signatures for moves
originally issued by agent $P$ and received at chain $A$.
\begin{property}
Here is the specification for the reliable delivery protocol.
\begin{itemize}
\item \emph{Authenticity}: Every move contained in a path signature
  in $\buffer_A(P)$ was signed by $P$.
\item \emph{Consistency}: If any receiver receives a path signature indicating $P$'s move,
  then, within $\Delta$, so does every other receiver.
\item \emph{Fairness}: If a compliant $P$ issues a move,
  then every receiver receives a path signature containing that move  within $\Delta$.
\end{itemize}
Note that a deviating sender may deliver multiple moves to the same receiver.
\end{property}

Before issuing a move,
a compliant agent $P$ waits until that move is enabled at some replica's state
(not shown in pseudocode).
To issue the move,
$P$ sends the path signature $[P](P,\mu,t)$ to every
replica (\algoref{front-end} \linref{fe:send}).
When replica $A$ receives a live path signature with a move
originally issued by $P$,
$A$ places that message in $\buffer_A(P)$
(\algoref{replica} \linref{rep:send}).
From that point, the relay service is in charge of delivery.

A natural way to structure the relay service
is to have each (compliant) sender run a dedicated thread that
repeatedly reads replica buffers and selectively relays messages
from one replica to the others (\algoref{relay}).
\linrefrange{relay:0}{relay:1} shows the pseudocode for relaying moves.
Each relaying agent $Q$ reads each receiver's buffer (\linref{relay:scan}),
and selects messages (more specifically, requests of moves) $(P, \mu,r)$ which are not already relayed by $Q$ (\linref{relay:live}).
Each such message is sent to the other receivers (\linref{relay:relay}) by adding $Q$ to the path $p$ and produces $[p,Q](P,\mu,r)$,
and the message is recorded to avoid later duplication (\linref{relay:seen}).
After reading the buffers,
$Q$ calls each replica's $\deliver()$ function (\linref{relay:deliver})
which causes the replica to check whether it can execute a move (see below).

\begin{theorem}
  Every move in $\buffer_A(P)$ was signed by $P$.
\end{theorem}

\begin{proof}
  Every move issued by $P$ is wrapped in a path signature signed by $P$
  (\algoref{front-end}, \linref{fe:send}).
\end{proof}

\begin{theorem}
  \thmlabel{delivered}
  If a path signature of the form $p(P,\mu,r)$ 
  is placed in any $\buffer_A(P)$ before $t+(r-1)n \Delta$,
  then a path signature for that request $(P,\mu,r)$
  will be placed in every $\buffer_B(P)$ before $t + r n \Delta$.
\end{theorem}

\begin{proof}
  The first time any replica $A$ places $p(P,\mu,r)$ in $\buffer_A(P)$
  where $p=[P_1,\ldots,P_k]$,
  then $p(P,\mu,r)$ must be live
  (\algoref{replica}, \linref{rep:live}).
  If that receipt occurred before $t + (r-1) n \Delta$,
  then liveness implies $t \leq k \Delta$.
  If any signer $P_i$ in $p$ was compliant,
  then $P_i$ already relayed that message to every replica.
  Otherwise, if no compliant agent has signed that message,
  then $k \leq n-1$, and within $\Delta$ some compliant agent $Q$
  will relay the path signature $[p,Q](P,\mu,r))$ to every replica.
  Each replica receives $[p, Q](P,\mu,r))$ before
  $t + \Delta + (r -1)n \Delta \leq (k+1) \Delta + (r -1)n \Delta$,
  so the path signature is live,
  and will be placed in $\buffer_B(P)$, for all $B \in \cASSET$.
\end{proof}

\begin{theorem}
  \thmlabel{compliant}
  If a compliant $P$ issues $\mu$ at round $r$,
  then a path signature containing $(P,\mu,r)$
  appears in $\buffer_A(P)$ for all $A \in \cASSET$ within time $\Delta$.
\end{theorem}

\begin{proof}
  Each compliant $P$ sends the path signature $[P](P,\mu,r)$ to each replica
  (\algoref{front-end}, \linref{fe:send}).
  That path signature is received by each replica $A$ before $\Delta$ has elapsed,
  so the path signature is live, and is placed in $\buffer_A(P)$
  (\algoref{replica}, \linref{rep:live}).
\end{proof}

\begin{algorithm}
  \caption{Relay Protocol for $Q \in \cA$}
  \algolabel{relay}
  \begin{algorithmic}[1]
    \State $\seen: 2^\cREQ$\Comment{Initially empty}
    \While{Exchange is in progress}
      \ForAll{$A \in \cASSET, P \in \cA$}\linlabel{relay:0}
        \ForAll{$p(P,\mu,r) \in \buffer_A(P)$}\Comment{Inspect every path signature}
        \linlabel{relay:scan}
          \If{$(P,\mu,r) \not\in \seen$}
              \Comment{Does it need to be relayed?}
              \linlabel{relay:live}
            \ForAll{$B \in \cASSET$}
              $B.\send([p,Q](P,\mu,r))$
              \Comment{Append signature and relay}
              \linlabel{relay:relay}
              \EndFor
            \State $\seen := \seen \cup \set{(P,\mu,r)}$\Comment{Don't relay again}
            \linlabel{relay:seen}
            \EndIf
          \EndFor
        \EndFor 
      \ForAll{$A \in \cASSET$}
        $A.\deliver()$\Comment{Wake up replicas}\linlabel{relay:1}
        \linlabel{relay:deliver}
        \EndFor
      \EndWhile
  \end{algorithmic}
\end{algorithm}
 
\subsection{Initialization, moves, and Settlement}
\sloppy Each replica manages a unique asset.
On the replica that manages asset $A$,
each agent $P$ has a \emph{long-lived account},
denoted $\longAccount_A(P)$,
that records how many units of $A$ are owned by $P$.
While an execution is in progress,
each agent $P$ has a \emph{short-lived account} at $A$,
denoted $\shortAccount_A(P,B)$,
tracking how many units of asset $B$ have been tentatively assigned to $P$ at replica $A$.
Each replica $A$ has address $\Self_A$.
Long-lived and short-lived accounts are related by the following invariant:
\begin{equation*}
\longAccount_A(\Self_A) = \sum_{P \in \cA}\shortAccount_A(P,A).
\end{equation*}
We assume replica $A$ is authorized to transfer $A$ assets, in either direction,
between the calling agent's long-lived account,
$\longAccount_A(\Sender)$,
and the replica's own long-lived account, $\longAccount_A(\Self)$.

At the start of the execution,
each agent $P$ escrows funds by transferring some quantity of
each asset $A$ from $\longAccount_A(P)$ to $\longAccount_A(\Self)$
(\algoref{replica}, \linref{rep:fund}).
If that transfer is successful,
the replica credits $P$'s short-lived accounts:
for all $B \in \cASSET$, $P$ sets $\shortAccount_B(P,A)$ equal to the amount funded
(\linref{rep:short-lived}).
Agent $P$ is then marked as funded (\linref{rep:funded}).
Only properly funded agents can execute moves (\linref{rep:live}).

What could go wrong?
The transfer from $\longAccount_A(P)$ to $\longAccount_A(\Self)$
might fail because $\longAccount_A(P)$ has insufficient funds.
The replica at $A$ can detect and react to such a failure,
but the other replicas cannot.
To protect against such failures,
each agent calls the function $\verifyAccounts()$ 
(\algoref{front-end}, \linrefrange{fe:acct:0}{fe:acct:1}),
which checks that all replicas' account balances are consistent.
Finally, each agent checks that every other agent has transferred
the agreed-upon amounts (\linref{fe:notok}).
This last test is application-specific:
for the swap example, agents would check that the others
transferred a specified amount of coins,
while in the DAO example,
LPs can transfer as many governance tokens as they like.
If either test fails, the front-end refunds that agent's assets (by invoking $redeem()$ in \algoref{front-end}).
In this way, if some agents drop out before initialization,
they are marked as unfunded,
and the remaining funded agents may or may not choose to continue.
In the meantime, \emph{safety} is preserved since each compliant agent who continues
sees a consistent state across all replicas.
Each compliant agent who leaves the execution gets their funds back,
ensuring they end up no worse off.

This funding step takes time at most $\Delta$.
Each agent then verifies that replicas are funded consistently.
If not, the agent calls \redeem() to reclaim its funding, and drops out.
This verification takes time $n\Delta$, like any other move.
For an execution starting at time $t$,
initialization completes before $t+(n+1)\Delta$.

While the execution is in progress,
transfers of asset $A$ between $P$ and $Q$ are expressed as transfers
between $\shortAccount_A(P,A)$ and $\shortAccount_A(Q,A)$,
leaving the balance of $\longAccount_A(\Self)$ unchanged.
Replica $A$ also tracks $P$'s balances for other assets:
$\shortAccount_A(P,B)$ is $A$'s view of $P$'s current short-lived balance
for each asset $B \neq A$.

\sloppy When the execution ends,
each $P$ calls each replica's $\redeem()$ function
(\algoref{front-end}, \linref{fe:redeem}),
to get its assets back.
This function transfers $\shortAccount_A(P,A)$ units of asset $A$
from $\longAccount_A(\Self)$ to $\longAccount_A(P)$
(\algoref{replica}, \linref{rep:redeem}).
Once an agent's funds are redeemed,
that agent is marked as not funded (\linref{rep:unfunded}).
The $\redeem()$ function serves two roles:
it can \emph{refund} an agent's original assets if the exchange fails,
or it can \emph{claim} an agent's new assets if the exchange succeeds. If all agents are conforming and no one drops out,
the execution proceeds and every agent ends up with a better payoff,
ensuring liveness.
Any compliant agent can drop out,
either early with a refund 
(say, if it observes inconsistent funding),
or at the execution's end.
Both choices ensure that all assets in short-lived accounts are moved to long-lived accounts.

\begin{algorithm}
  \caption{Replica for asset $A$}
  \algolabel{replica}
  \begin{algorithmic}[1]
    \Component
    \State $\buffer_A: \cA \to 2^{\cPS}$ \Comment{initially all $\emptyset$}
    \State $\funded_A: \cA \to \set{\true, \false}$ \Comment{initially all $\false$}
    \State $\machine$ \Comment{Replica of state machine}\linlabel{rep:machine}
    \State $\delivered_A: 2^\cREQ$\Comment{Next move to execute}
     \State $\startTime:\mathbb{Z} \to  \mathbb{Z}$\Comment{Each round's start time}
    \Initial{$\fund: \cASSET \to \mathbb{Z}$}\Comment{Initial funding}
      \State transfer $\fund(A)$ units of $A$
         from $\longAccount_A(\Sender)$ to $\longAccount_A(\Self)$\linlabel{rep:fund}
      \If {transfer was successful}
        \ForAll{$B \in \cASSET$}
          $\shortAccount_B(\Sender,A) := \fund(A)$\linlabel{rep:short-lived}
          \EndFor
        \State $\funded_A(\Sender) := \true$\linlabel{rep:funded}
        \EndIf
      \State $\startTime(1) := (n+1)\Delta$ \Comment{Set start time for round 1}
    \Move send $p(P, \mu, t)$\Comment{Forward move to relay protocol}
      \If{$\live(p(P, \mu, t)$ and $\funded_A(P)$}\linlabel{rep:live}
        \Comment{Relay only live moves by funded agents}
        \State $\buffer_A(P) := \buffer_A(P) \cup \set{p(P,\mu,r)}$
        \linlabel{rep:send}
        \EndIf
    \Move $\deliver()$\Comment{Relayer: check for executable move}
      \State $\delivered := 
      \{(P,\mu,r) \;|\; p(P,\mu,r) \in \buffer_A(P) \wedge \ready(p(P, \mu, r))$ $\wedge\; P \in \enabled(\machine) \wedge \mu \in \moves(\machine)\}$
        \Comment{Execute if enough time has elapsed}
        \linlabel{rep:enabled}
      \If{$\delivered = \set{(P, \mu, t)}$ }\Comment{Unique executable move?}
        \State $\machine.\mu()$\linlabel{rep:unique}\Comment{Execute it}
        \State $\startTime(r+1) :=\startTime(r)+ n \Delta $ \Comment{Set the start time of next round}\linlabel{rep:timer}
      \ElsIf{$\now()-\startTime(r) > n \Delta$} \Comment{Did we time out?}
        \linlabel{rep:elapse}
        \State $\machine.\skipMove()$\linlabel{rep:skip}\Comment{Agent chose not to move}
        \State $\startTime(r+1) :=\startTime(r)+ n \Delta$\Comment{Set the start time of next round}
      \EndIf

    \Move $\topUp(\fund: \cASSET \to \mathbb{Z})$\Comment{Dynamically add funding}
      \If{$\funded_A(\Sender)$}\linlabel{rep:topup:0}
        \State transfer $\fund(A)$ units of $A$
           from $\longAccount_A(\Sender)$ to $\longAccount_A(\Self)$\linlabel{rep:topup:fund}
        \If {transfer was successful}
          \State $(\forall B \in \cASSET) \shortAccount_B(\Sender,A) := \shortAccount_B(\Sender,A) + \fund(A)$\Comment{Credit accounts}\linlabel{rep:topup:short-lived}
        \Else
          \State $\funded_A(\Sender) := \false$\Comment{Freeze accounts}\linlabel{rep:topup:1}
          \EndIf
        \EndIf
    \Move $\defund(\defundVote: \cA \to \set{\true,\false})$\linlabel{rep:defund:0}
     \If{$\Sender = \leader$}\Comment{check authorization}
       \ForAll{$P \in \cA$}
         \If{$\defundVote(P)$}
           \State $\funded_A(P) := \false$\linlabel{rep:defund:1}
           \EndIf
         \EndFor
       \EndIf
      \Move $redeem()$\Comment{Settle accounts at end}
        \If{$\funded_A(\Sender)$}
          \State transfer $\shortAccount_A(A,\Sender)$ units of $A$
            from $\longAccount_A(\Self)$ to $\longAccount_A(\Sender)$\linlabel{rep:redeem}
          \State $\funded_A(\Sender) := \false$\linlabel{rep:unfunded}
          \Comment{No moves allowed after cashing out}
          \EndIf
  \end{algorithmic}
\end{algorithm}

Each replica (\algoref{replica}) has its own copy of the state machine
(\linref{rep:machine}).
The replica can apply moves to the copy (\linref{rep:unique}),
and the replica can determine whether a proposed move
by a particular agent is currently enabled (\linref{rep:enabled}).

Replica $A$'s $\deliver()$ function determines
whether there is a unique move in $\buffer_A(P)$ to execute.
Each time replica $A$ starts a round,
it records the time (\linref{rep:timer}).
If $n\Delta$ time then elapses without delivering
the next move (\linref{rep:elapse}),
the missing move is deemed to be a $\skipMove$ (\linref{rep:skip}).

\subsection{Dynamic Funding}
In the auction state machine (\gameref{auction}),
each agent bids without knowing the others' bids,
but each agent does know the others' maximum possible bids,
because short-lived account balances are public.
For example,
if Alice initially funded her short-lived account with 200 coins and Bob with 100,
then Alice knows she can win by bidding 101 coins.

What if agents could provide additional funding while the execution is in progress?
For example, an English auction with multiple bidding rounds
might pause between rounds to allow agents to
add more coins to their short-lived accounts.
If Alice bids 101 coins in the first round,
Bob might respond by adding enough coins to his short-lived account
to respond in the next round.

Such flexibility, however, comes at a cost.
Consider two scenarios.
In the first,
Bob transfers a million coins from his long-lived account
to his short-lived account.
This transfer happens on the coin replica.
When he honestly reports that transfer to the auction replica, 
Alice and Carol consistently and falsely report to the auction replica
that the transfer never occurred.
In the second scenario,
Bob fails to transfer the million coins for lack of funds.
When he dishonestly reports a successful transfer to the auction replica, 
Alice and Carol consistently and honestly report the transfer never occurred.

An ideal mechanism would ensure that in the first scenario,
where Bob's transfer succeeded on the coin replica,
his claim would be accepted on the auction replica,
and the auction would continue with Bob's new funding.
In the second scenario, however,
where Bob's transfer failed on the coin replica,
his claim would be rejected on the auction replica,
and the auction would continue without him.
Unfortunately,
these scenarios are indistinguishable to the auction replica,
which cannot directly observe the transfer's outcome on the coin replica.
Alice and Carol know whether Bob is lying,
but they have no way to ``convince'' the auction replica.

An imperfect way to support dynamic funding is to make it expensive for Bob to cheat.
As part of initialization, each agent pays a deposit at each replica.
After each round of bidding,
the bidders execute a \emph{top-up} round,
similar to the initialization round.
Each bidder may transfer additional coins from his long-lived account to
the state machine's coin account, a sum reflected in his short-lived account.
If, for example, Bob's transfer fails,
he forfeits his deposit, along with any coins transferred in earlier rounds.
At the end of the top-up round,
the bidders call $\verifyAccounts()$,
withdrawing and quitting if accounts are incorrect.
This mechanism is imperfect because it gives
Bob the power to wreck the auction through a (perhaps deliberately) failed transfer,
but if he does so, he pays a penalty.

\algoref{front-end}, \linrefrange{fe:topup:0}{fe:topup:1} shows pseudocode
for the front-end's $\topUp()$ function that dynamically adds funding to an execution.
Just as for initialization,
each agent specifies, for each asset $A$,
how many additional units of $A$ to transfer to the execution (\linref{fe:topup:send}).
Each agent then checks that the new global funding state is consistent,
and if not, it redeems and quits (\linref{fe:topup:1}).

\algoref{replica}, \linrefrange{rep:topup:0}{rep:topup:1} shows pseudocode
for replica $A$'s $\topUp()$ function.
The function transfers the additional units of $A$ from to the replica's account
(\linref{rep:topup:fund}).
If the transfer is successful, the short-lived accounts are credited
(\linref{rep:topup:short-lived}).
If the transfer is not successful,
the agent is marked as unfunded (\linref{rep:topup:1}),
causing the caller to forfeit his deposit, along with any previous asset transfers.
  
A second imperfect way to support dynamic funding is to
limit Bob's potential for mischief by giving Carol, the auctioneer,
authority to expel bidders whose transfers fail.
(Carol's role could also be filled by a committee.)
This mechanism is imperfect because it grants Carol
the power to expel Bob under false pretexts,
although she has no incentive to do so,
since higher bids mean higher profits for her.
Note that Bob does not lose any assets if he is unfairly expelled,
so he ends up no worse off.

At initialization, Carol is installed as the \emph{leader} at each replica (not shown).
After each round of bidding,
the bidders execute a \emph{verified top-up} round
(\algoref{front-end}, \linrefrange{fe:ver:0}{fe:ver:1}) as
an extension of the top-up round described earlier.
Each agent executes a top-up round as before (\linref{fe:ver:0}).
The leader (Carol) then inspects the accounts,
records which agents' transfers failed (\linref{fe:ver:1}),
and instructs the replicas to mark those accounts as defunded
(\linref{fe:ver:defund} and \algoref{replica}, \linrefrange{rep:defund:0}{rep:defund:1}),
ensuring they will henceforth be ignored by the rest of the agents.
Finally, all agents call $\verifyAccounts()$ 
before moving on to the next round of bidding.

\begin{algorithm}[htb]
  \caption{Front-end for agent $P$}
  \algolabel{front-end}
  \begin{algorithmic}[1]
    \Initial{$\fund_P: \cASSET \to \mathbb{Z}$}
      \ForAll{$A \in \cASSET$}
        $A.\initial(\fund_P)$\linlabel{fe:fund}
        \EndFor
      \State $\verifyAccounts()$
      \If{$(\exists A \in \cASSET, P \in \cA) \shortAccount_A(P,A)$ is not the agreed-upon amount}
      \linlabel{fe:notok}
        \State $\redeem()$\Comment{Get assets refunded}
        \State \halt
        \EndIf
    \Move $\verifyAccounts()$\Comment{Check for malformed funding data}
      \If{$(\exists A,B \in \cASSET, Q \in \cA)\; \funded_A(Q) \wedge (\shortAccount_A(Q,A) \neq \shortAccount_B(Q,A))$}
        \linlabel{fe:acct:0}
        \State $\redeem()$\Comment{Ask for asset refund}
        \State \halt\linlabel{fe:acct:1}
        \EndIf
    \Move send$(\mu)$
      \ForAll{$A \in \cASSET$}
        \State $A.send([P](P, \mu, t))$\Comment{Add path sig and broadcast}
        \linlabel{fe:send}
        \EndFor
    \Move $\topUp(\fund_P: \cASSET \to \mathbb{Z})$\linlabel{fe:topup:0}
      \ForAll{$A \in \cASSET$}
        $A.\topUp(\fund_P)$\Comment{Send path sig}\linlabel{fe:topup:send}
        \EndFor
      \State $\verifyAccounts()$\linlabel{fe:topup:1}
    \Move $\topUpVerified(\fund: \cASSET \to \mathbb{Z})$
      \State $\topUp(fund_p)$\Comment{Call regular top-up}\linlabel{fe:ver:0}
      \If{$\Self = \leader$}\Comment{Authorized to accept or reject top-up}
        \ForAll{$P \in \cA$}
          \State $\defundVote(P) :=
\sloppy            (\exists A,B \in \cASSET) \shortAccount_A(P,A) \neq \shortAccount_B(P,A)$\linlabel{fe:ver:detect}
          \EndFor
        \ForAll{$A \in \cASSET$}
          $A.\defund(\defundVote)$\linlabel{fe:ver:defund}
          \EndFor
      \Else
        \State wait for leader to deliver defund votes
        \EndIf
      \State $\verifyAccounts()$\linlabel{fe:ver:1}
    \Move $\redeem()$\Comment{Reclaim assets}\linlabel{fe:redeem}
        \ForAll{$A \in \cASSET$}
          $A.\redeem()$\Comment{Redeem assets from each replica}
          \EndFor
  \end{algorithmic}
\end{algorithm}

\section{Remarks}
\seclabel{remarks}
Although it is mostly straightforward to implement replica automata
as smart contracts,
there are practical, blockchain-specific details (such as analyzing gas prices)
that are beyond the scope of this paper.

We model abstract state machines in adversarial environments as sequential games,
where agents take turns.
In some applications, agents can make moves in parallel,
as long as certain conditions hold.
In the simple swap example, Alice and Bob can agree in parallel.
In the DAO example,
votes can be cast in parallel,
and in the auction example,
sealed bids can be submitted in parallel,
bids can be unsealed in parallel,
and outcomes can be resolved in parallel.

When is it safe to execute moves in parallel?
First, parallel moves must \emph{commute},
meaning that the moves' order does not matter.
Commutativity is essential because thos moves may appear in
different orders at different replicas.
Second, the moves should be \emph{strategically independent},
meaning that no agent would change its move if it were
aware of a parallel move.
For example, Alice and Bob should not issue concurrent plain-text bids,
because a corrupt validator might leak Alice's bid to Bob,
and allow Bob to change his bid in response.
By contrast, Alice and Bob are free to issue sealed bids in parallel,
because neither agent could benefit from observing the other's sealed bid.

The principal cost of our SMR protocol is that emulating a state machine
transition requires time $n\Delta$.
One way to reduce this cost is to proceed \emph{optimistically}:
each replica executes each enabled move as soon as it is received,
and overlaps subsequent execution with waiting to detect duplicates.
When the execution completes,
if there are still moves waiting to be finalized,
the protocol simply postpones termination until the absence of conflicts is confirmed.
If a conflict is detected,
then the execution must be rolled back, either to the point of conflict,
or to initialization, and retried, perhaps without speculation.
Speculation transforms an $r$-move execution
from time $O(rN) \Delta$ to time $O(r+N)\Delta$, for failure-free executions.

Recall \gameref{swap},
where Alice trades her florin for Bob's ducat.
If Alice's final move (\linref{swap:alice}) is delivered
to the ducat replica but not the florin replica,
then Alice collects both coins.
This injustice cannot happen as long as Bob is compliant,
but what if Bob's machine crashes,
or he is the victim of a denial-of-service attack?
Technically, Bob would be at fault,
but he could protect himself by enlisting additional agents,
who do not appear in the state machine specification, simply to relay messages.
(A similar observation applies to the Lightning payment network~\cite{lightning},
where one can hire a \emph{watchtower} service~\cite{watchtower}
to act on the behalf of an agent that might inadvertently go off-line.)

How does the SMR protocol compare to specialized protocols?
In compliant executions,
the Nolan two-agent swap~\cite{tiersnolan} completes in time at most $4\Delta$.
\gameref{swap} requires $2\Delta$ to initialize,
$2\Delta$ for each agreement (which could be done in parallel),
and $2\Delta$ to complete the transfers.
An ad-hoc multi-agent swap protocol~\cite{Herlihy2018} has $O(n\Delta)$ latency,
as does a comparable SMR protocol.
Latency is harder to compare for executions in which agents deviate,
since the SMR protocol may succeed in some scenarios where the \emph{ad-protocols} fail.

\seclabel{generic}
Our generic SMR protocol may be useful in other contexts.
Recall that we replicate computation as well as data because one contract
cannot directly observe other's state.
Suppose instead that there is one \emph{main} chain whose contracts
are capable of producing \emph{proofs} of its current state,
and that these proofs can be checked by contracts at the other chains.
(Examples in the literature include \emph{ZK rollups}~\cite{buterin2021},
and the \emph{certifying blockchain} of Herlihy \emph{et al.}~\cite{HerlihyLS2021}
used to commit atomic cross-chain transactions.)
The execution would take place only on the main chain,
while the other chains need only track the state by checking proofs,
using the generic SMR protocol to ensure that
emphemeral accounts are replicated consistently,
so assets are distributed correctly at the end of the execution.

Instead of producing a proof,
the main chain could produce an \emph{optimistic rollup}~\cite{buterin2021,Kalodner2018},
a summary listing of the agents' final ephemeral account balances.
This summary is replicated via the generic SMR protocol at the chains,
but asset distribution is delayed long enough to allow agents time to detect fraud,
and if found, to publish a proof and collect a reward.

In these contexts, our SMR protocol serves as a trustless \emph{token bridge}~\cite{Sidhu2020},
effectively transferring tokens or coins from each replica chain to the main chain
at the execution's start,
then transferring them back (to their new owners) at the end.

\section{Related Work}
\seclabel{related}
State machine replication is a classic problem in distributed computing.
Protocols such as Paxos~\cite{Lamport1998}, Raft~\cite{OngaroO2014},
and their immediate descendants were designed to tolerate crash failures.
Later protocols (see Distler's survey~\cite{distler2021}) tolerate Byzantine failures.
As noted,
these protocols are not applicable to cross-chain exchanges
because of differences in the underlying trust and communication models.

Prior Byzantine fault-tolerant (BFT) SMR protocols ~\cite{distler2021} assume
replicas may be Byzantine but clients are honest.
By contrast, in our cross-chain model,
replicas are correct (because they represent blockchains),
but clients can be Byzantine (because they may try to steal one another's assets).

Except for some randomized protocols\footnote{The following SMR protocols do not have explicit leaders \cite{moniz2008ritas,duan2018beat,miller2016}.},
most prior BFT-SMR protocols assign one replica to be the \emph{leader},
and the rest to be \emph{followers} (sometimes ``validators'').
These protocols tolerate only a certain fraction of faulty replicas.
Our cross-chain SMR protocol, by contrast,
does not distinguish between leaders and followers,
and tolerates any number of faulty agents.

An individual blockchain's consensus protocol can be viewed as an SMR protocol,
where the ledger state is replicated among the validators (miners).
Validators are typically rewarded for
participating~\cite{amoussouguenou,bitcoin,roughgarden2020transaction}. Validators might deviate in various ways,
including \emph{selfish mining}~\cite{eyal2014selfishMining},
front-running~\cite{daian2019flash},
or exploiting the structure of consensus rewards~\cite{abraham2018, amoussou2020rational, amoussou2020rational2, bano2019sok, buterin2020incentives, daian2019snow,kiayias2017ouroboros,kothapalli2017smartcast,pass2017fruitchains,rocsu2021evolution,saleh2021blockchain,sliwinski2020blockchains}.
Individual blockchain SMR protocols are not applicable for cross-chain SMR
because of fundamental differences in models and participants' incentives.

An alternative approach to cross-chain interoperability is allowing blockchains to
communicate states to one another.
As pointed out by a recent survey \cite{qasse2019inter},
most such solutions work only for homogeneous blockchains.
These protocols usually adopt external relayers/validators to relay/validate messages across chains.
Any failure of those external players can harm the safety of the cross-chain system.
Incentivizing external operators remains a challenge~\cite{robinson2021survey}.

In failure models of some prior work, parties are classified as either rational,
seeking to maximize payoffs,
or Byzantine, capable of any behavior.
First proposed for distributed systems~\cite{moscibroda2006selfish},
this classification has been used for chain consensus
protocols~\cite{lev2019fairledger,sliwinski2020blockchains,mcmenamin2021achieving}.
The rational vs Byzantine classification is equivalent to our compliant vs deviating
classification for cross-chain exchanges where compliance is rational,
a property one would expect in practice.

The notion of replacing an \emph{ad-hoc} protocol with a generic,
replicated state machine was anticipated by Miller \emph{et al.}~\cite{MillerBKM17},
who propose generic \emph{state channels} as a cleaner, systematic
replacement for prior payment channels of the kind used in the Lightning network.

\newpage

\bibliographystyle{ACM-Reference-Format}
\bibliography{zotero,references}  


\begin{thebibliography}{45}


\ifx \showCODEN    \undefined \def \showCODEN     #1{\unskip}     \fi
\ifx \showDOI      \undefined \def \showDOI       #1{#1}\fi
\ifx \showISBNx    \undefined \def \showISBNx     #1{\unskip}     \fi
\ifx \showISBNxiii \undefined \def \showISBNxiii  #1{\unskip}     \fi
\ifx \showISSN     \undefined \def \showISSN      #1{\unskip}     \fi
\ifx \showLCCN     \undefined \def \showLCCN      #1{\unskip}     \fi
\ifx \shownote     \undefined \def \shownote      #1{#1}          \fi
\ifx \showarticletitle \undefined \def \showarticletitle #1{#1}   \fi
\ifx \showURL      \undefined \def \showURL       {\relax}        \fi
\providecommand\bibfield[2]{#2}
\providecommand\bibinfo[2]{#2}
\providecommand\natexlab[1]{#1}
\providecommand\showeprint[2][]{arXiv:#2}

\bibitem[\protect\citeauthoryear{Abraham, Malkhi, Nayak, Ren, and
  Spiegelman}{Abraham et~al\mbox{.}}{2017}]%
        {abraham2018}
\bibfield{author}{\bibinfo{person}{Ittai Abraham}, \bibinfo{person}{Dahlia
  Malkhi}, \bibinfo{person}{Kartik Nayak}, \bibinfo{person}{Ling Ren}, {and}
  \bibinfo{person}{Alexander Spiegelman}.} \bibinfo{year}{2017}\natexlab{}.
\newblock \showarticletitle{Solida: {A} {Blockchain} {Protocol} {Based} on
  {Reconfigurable} {Byzantine} {Consensus}}.
\newblock \bibinfo{journal}{\emph{arXiv:1612.02916 [cs]}} (\bibinfo{date}{Nov.}
  \bibinfo{year}{2017}).
\newblock
\urldef\tempurl%
\url{http://arxiv.org/abs/1612.02916}
\showURL{%
\tempurl}
\newblock
\shownote{arXiv: 1612.02916.}


\bibitem[\protect\citeauthoryear{Abraham, Malkhi, Nayak, Ren, and Yin}{Abraham
  et~al\mbox{.}}{2020}]%
        {synchot}
\bibfield{author}{\bibinfo{person}{Ittai Abraham}, \bibinfo{person}{Dahlia
  Malkhi}, \bibinfo{person}{Kartik Nayak}, \bibinfo{person}{Ling Ren}, {and}
  \bibinfo{person}{Maofan Yin}.} \bibinfo{year}{2020}\natexlab{}.
\newblock \showarticletitle{Sync HotStuff: Simple and Practical Synchronous
  State Machine Replication}. In \bibinfo{booktitle}{\emph{2020 IEEE Symposium
  on Security and Privacy (SP)}}. \bibinfo{pages}{106--118}.
\newblock
\urldef\tempurl%
\url{https://doi.org/10.1109/SP40000.2020.00044}
\showDOI{\tempurl}


\bibitem[\protect\citeauthoryear{Amoussou-Guenou, Biais, Potop-Butucaru, and
  Tucci-Piergiovanni}{Amoussou-Guenou et~al\mbox{.}}{2019a}]%
        {amoussou2020rational}
\bibfield{author}{\bibinfo{person}{Yackolley Amoussou-Guenou},
  \bibinfo{person}{Bruno Biais}, \bibinfo{person}{Maria Potop-Butucaru}, {and}
  \bibinfo{person}{Sara Tucci-Piergiovanni}.} \bibinfo{year}{2019}\natexlab{a}.
\newblock \showarticletitle{Rationals vs {Byzantines} in {Consensus}-based
  {Blockchains}}.
\newblock \bibinfo{journal}{\emph{arXiv:1902.07895 [cs]}} (\bibinfo{date}{Feb.}
  \bibinfo{year}{2019}).
\newblock
\urldef\tempurl%
\url{http://arxiv.org/abs/1902.07895}
\showURL{%
\tempurl}
\newblock
\shownote{arXiv: 1902.07895.}


\bibitem[\protect\citeauthoryear{Amoussou-Guenou, Biais, Potop-Butucaru, and
  Tucci-Piergiovanni}{Amoussou-Guenou et~al\mbox{.}}{2020}]%
        {amoussou2020rational2}
\bibfield{author}{\bibinfo{person}{Yackolley Amoussou-Guenou},
  \bibinfo{person}{Bruno Biais}, \bibinfo{person}{Maria Potop-Butucaru}, {and}
  \bibinfo{person}{Sara Tucci-Piergiovanni}.} \bibinfo{year}{2020}\natexlab{}.
\newblock \bibinfo{title}{Rational Behavior in Committee-Based Blockchains}.
\newblock \bibinfo{howpublished}{Cryptology ePrint Archive, Report 2020/710}.
\newblock
\newblock
\shownote{\url{https://ia.cr/2020/710}.}


\bibitem[\protect\citeauthoryear{Amoussou-Guenou, del Pozzo, Potop-Butucaru,
  and Tucci-Piergiovanni}{Amoussou-Guenou et~al\mbox{.}}{2019b}]%
        {amoussouguenou}
\bibfield{author}{\bibinfo{person}{Yackolley Amoussou-Guenou},
  \bibinfo{person}{Antonella del Pozzo}, \bibinfo{person}{Maria
  Potop-Butucaru}, {and} \bibinfo{person}{Sara Tucci-Piergiovanni}.}
  \bibinfo{year}{2019}\natexlab{b}.
\newblock \showarticletitle{On {Fairness} in {Committee}-based {Blockchains}}.
\newblock \bibinfo{journal}{\emph{arXiv:1910.09786 [cs]}} (\bibinfo{date}{Oct.}
  \bibinfo{year}{2019}).
\newblock
\urldef\tempurl%
\url{http://arxiv.org/abs/1910.09786}
\showURL{%
\tempurl}
\newblock
\shownote{arXiv: 1910.09786.}


\bibitem[\protect\citeauthoryear{Bano, Sonnino, Al-Bassam, Azouvi, McCorry,
  Meiklejohn, and Danezis}{Bano et~al\mbox{.}}{2017}]%
        {bano2019sok}
\bibfield{author}{\bibinfo{person}{Shehar Bano}, \bibinfo{person}{Alberto
  Sonnino}, \bibinfo{person}{Mustafa Al-Bassam}, \bibinfo{person}{Sarah
  Azouvi}, \bibinfo{person}{Patrick McCorry}, \bibinfo{person}{Sarah
  Meiklejohn}, {and} \bibinfo{person}{George Danezis}.}
  \bibinfo{year}{2017}\natexlab{}.
\newblock \showarticletitle{Consensus in the {Age} of {Blockchains}}.
\newblock \bibinfo{journal}{\emph{arXiv:1711.03936 [cs]}} (\bibinfo{date}{Nov.}
  \bibinfo{year}{2017}).
\newblock
\urldef\tempurl%
\url{http://arxiv.org/abs/1711.03936}
\showURL{%
\tempurl}
\newblock
\shownote{arXiv: 1711.03936.}


\bibitem[\protect\citeauthoryear{Bracha}{Bracha}{1987}]%
        {Bracha1987}
\bibfield{author}{\bibinfo{person}{Gabriel Bracha}.}
  \bibinfo{year}{1987}\natexlab{}.
\newblock \showarticletitle{Asynchronous {Byzantine} agreement protocols}.
\newblock \bibinfo{journal}{\emph{Information and Computation}}
  \bibinfo{volume}{75}, \bibinfo{number}{2} (\bibinfo{date}{Nov.}
  \bibinfo{year}{1987}), \bibinfo{pages}{130--143}.
\newblock
\showISSN{08905401}
\urldef\tempurl%
\url{https://doi.org/10.1016/0890-5401(87)90054-X}
\showDOI{\tempurl}


\bibitem[\protect\citeauthoryear{Buterin, Reijsbergen, Leonardos, and
  Piliouras}{Buterin et~al\mbox{.}}{2020}]%
        {buterin2020incentives}
\bibfield{author}{\bibinfo{person}{Vitalik Buterin}, \bibinfo{person}{Daniël
  Reijsbergen}, \bibinfo{person}{Stefanos Leonardos}, {and}
  \bibinfo{person}{Georgios Piliouras}.} \bibinfo{year}{2020}\natexlab{}.
\newblock \showarticletitle{Incentives in {Ethereum}'s hybrid {Casper}
  protocol}.
\newblock \bibinfo{journal}{\emph{International Journal of Network Management}}
  \bibinfo{volume}{30}, \bibinfo{number}{5} (\bibinfo{date}{Sept.}
  \bibinfo{year}{2020}).
\newblock
\showISSN{1055-7148, 1099-1190}
\urldef\tempurl%
\url{https://doi.org/10.1002/nem.2098}
\showDOI{\tempurl}


\bibitem[\protect\citeauthoryear{Castro and Liskov}{Castro and Liskov}{1999}]%
        {PBFT}
\bibfield{author}{\bibinfo{person}{Miguel Castro} {and}
  \bibinfo{person}{Barbara Liskov}.} \bibinfo{year}{1999}\natexlab{}.
\newblock \showarticletitle{Practical byzantine fault tolerance}. In
  \bibinfo{booktitle}{\emph{Proceedings of the third symposium on operating
  systems design and implementation}} \emph{(\bibinfo{series}{{OSDI} '99})}.
  \bibinfo{publisher}{USENIX Association}, \bibinfo{address}{Berkeley, CA,
  USA}, \bibinfo{pages}{173--186}.
\newblock
\showISBNx{1-880446-39-1}
\urldef\tempurl%
\url{http://dl.acm.org/citation.cfm?id=296806.296824}
\showURL{%
\tempurl}
\newblock
\shownote{Number of pages: 14 Place: New Orleans, Louisiana, USA tex.acmid:
  296824.}


\bibitem[\protect\citeauthoryear{Chester}{Chester}{2018}]%
        {watchtower}
\bibfield{author}{\bibinfo{person}{Jonathan Chester}.}
  \bibinfo{year}{2018}\natexlab{}.
\newblock \bibinfo{title}{Your guide on bitcoin's lightning network: {The}
  opportunities and the issues}.
\newblock
\newblock
\urldef\tempurl%
\url{https://www.forbes.com/sites/jonathanchester/2018/06/18/your-guide-on-the-lightning-network-the-opportunities-and-the-issues/#6c8d8c0f3677N}
\showURL{%
\tempurl}


\bibitem[\protect\citeauthoryear{Daian, Goldfeder, Kell, Li, Zhao, Bentov,
  Breidenbach, and Juels}{Daian et~al\mbox{.}}{2019a}]%
        {daian2019flash}
\bibfield{author}{\bibinfo{person}{Philip Daian}, \bibinfo{person}{Steven
  Goldfeder}, \bibinfo{person}{Tyler Kell}, \bibinfo{person}{Yunqi Li},
  \bibinfo{person}{Xueyuan Zhao}, \bibinfo{person}{Iddo Bentov},
  \bibinfo{person}{Lorenz Breidenbach}, {and} \bibinfo{person}{Ari Juels}.}
  \bibinfo{year}{2019}\natexlab{a}.
\newblock \showarticletitle{Flash {Boys} 2.0: {Frontrunning}, {Transaction}
  {Reordering}, and {Consensus} {Instability} in {Decentralized} {Exchanges}}.
\newblock \bibinfo{journal}{\emph{arXiv:1904.05234 [cs]}}
  (\bibinfo{date}{April} \bibinfo{year}{2019}).
\newblock
\urldef\tempurl%
\url{http://arxiv.org/abs/1904.05234}
\showURL{%
\tempurl}
\newblock
\shownote{arXiv: 1904.05234.}


\bibitem[\protect\citeauthoryear{Daian, Pass, and Shi}{Daian
  et~al\mbox{.}}{2019b}]%
        {daian2019snow}
\bibfield{author}{\bibinfo{person}{Phil Daian}, \bibinfo{person}{Rafael Pass},
  {and} \bibinfo{person}{Elaine Shi}.} \bibinfo{year}{2019}\natexlab{b}.
\newblock \showarticletitle{Snow {White}: {Robustly} {Reconfigurable}
  {Consensus} and {Applications} to {Provably} {Secure} {Proof} of {Stake}}.
\newblock In \bibinfo{booktitle}{\emph{Financial {Cryptography} and {Data}
  {Security}}}, \bibfield{editor}{\bibinfo{person}{Ian Goldberg} {and}
  \bibinfo{person}{Tyler Moore}} (Eds.). Vol.~\bibinfo{volume}{11598}.
  \bibinfo{publisher}{Springer International Publishing},
  \bibinfo{address}{Cham}, \bibinfo{pages}{23--41}.
\newblock
\showISBNx{978-3-030-32100-0 978-3-030-32101-7}
\urldef\tempurl%
\url{https://doi.org/10.1007/978-3-030-32101-7_2}
\showDOI{\tempurl}
\newblock
\shownote{Series Title: Lecture Notes in Computer Science.}


\bibitem[\protect\citeauthoryear{{Diego Ongaro and John Ousterhout}}{{Diego
  Ongaro and John Ousterhout}}{2014}]%
        {OngaroO2014}
\bibfield{author}{\bibinfo{person}{{Diego Ongaro and John Ousterhout}}.}
  \bibinfo{year}{2014}\natexlab{}.
\newblock \showarticletitle{In {Search} of an {Understandable} {Consensus}
  {Algorithm}}. In \bibinfo{booktitle}{\emph{2014 {USENIX} {Annual} {Technical}
  {Conference}}}. \bibinfo{publisher}{USENIX Association},
  \bibinfo{address}{Philadelphia, PA}, \bibinfo{pages}{305--319}.
\newblock
\showISBNx{978-1-931971-10-2}
\urldef\tempurl%
\url{https://www.usenix.org/conference/atc14/technical-sessions/presentation/ongaro}
\showURL{%
\tempurl}


\bibitem[\protect\citeauthoryear{Distler}{Distler}{2021}]%
        {distler2021}
\bibfield{author}{\bibinfo{person}{Tobias Distler}.}
  \bibinfo{year}{2021}\natexlab{}.
\newblock \showarticletitle{Byzantine {Fault}-tolerant {State}-machine
  {Replication} from a {Systems} {Perspective}}.
\newblock \bibinfo{journal}{\emph{Comput. Surveys}} \bibinfo{volume}{54},
  \bibinfo{number}{1} (\bibinfo{date}{April} \bibinfo{year}{2021}),
  \bibinfo{pages}{1--38}.
\newblock
\showISSN{0360-0300, 1557-7341}
\urldef\tempurl%
\url{https://doi.org/10.1145/3436728}
\showDOI{\tempurl}


\bibitem[\protect\citeauthoryear{Duan, Reiter, and Zhang}{Duan
  et~al\mbox{.}}{2018}]%
        {duan2018beat}
\bibfield{author}{\bibinfo{person}{Sisi Duan}, \bibinfo{person}{Michael~K
  Reiter}, {and} \bibinfo{person}{Haibin Zhang}.}
  \bibinfo{year}{2018}\natexlab{}.
\newblock \showarticletitle{BEAT: Asynchronous BFT made practical}. In
  \bibinfo{booktitle}{\emph{Proceedings of the 2018 ACM SIGSAC Conference on
  Computer and Communications Security}}. \bibinfo{pages}{2028--2041}.
\newblock
\urldef\tempurl%
\url{https://doi.org/10.1145/3436728}
\showURL{%
\tempurl}


\bibitem[\protect\citeauthoryear{Eyal and Sirer}{Eyal and Sirer}{2014}]%
        {eyal2014selfishMining}
\bibfield{author}{\bibinfo{person}{Ittay Eyal} {and} \bibinfo{person}{Emin~Gün
  Sirer}.} \bibinfo{year}{2014}\natexlab{}.
\newblock \showarticletitle{Majority is not enough: {Bitcoin} mining is
  vulnerable}. In \bibinfo{booktitle}{\emph{Financial cryptography and data
  security - 18th international conference, {FC} 2014, christ church, barbados,
  march 3-7, 2014, revised selected papers}}. \bibinfo{pages}{436--454}.
\newblock
\urldef\tempurl%
\url{https://doi.org/10.1007/978-3-662-45472-5_28}
\showDOI{\tempurl}
\newblock
\shownote{tex.bibsource: dblp computer science bibliography, http://dblp.org
  tex.biburl: http://dblp.uni-trier.de/rec/bib/conf/fc/EyalS14 tex.timestamp:
  Mon, 10 Nov 2014 13:58:26 +0100.}


\bibitem[\protect\citeauthoryear{Garay, Kiayias, and Leonardos}{Garay
  et~al\mbox{.}}{2015}]%
        {GarayKL2015}
\bibfield{author}{\bibinfo{person}{Juan Garay}, \bibinfo{person}{Aggelos
  Kiayias}, {and} \bibinfo{person}{Nikos Leonardos}.}
  \bibinfo{year}{2015}\natexlab{}.
\newblock \showarticletitle{The {Bitcoin} {Backbone} {Protocol}: {Analysis} and
  {Applications}}.
\newblock In \bibinfo{booktitle}{\emph{Advances in {Cryptology} - {EUROCRYPT}
  2015}}, \bibfield{editor}{\bibinfo{person}{Elisabeth Oswald} {and}
  \bibinfo{person}{Marc Fischlin}} (Eds.). Vol.~\bibinfo{volume}{9057}.
  \bibinfo{publisher}{Springer Berlin Heidelberg}, \bibinfo{address}{Berlin,
  Heidelberg}, \bibinfo{pages}{281--310}.
\newblock
\showISBNx{978-3-662-46802-9 978-3-662-46803-6}
\urldef\tempurl%
\url{https://doi.org/10.1007/978-3-662-46803-6_10}
\showDOI{\tempurl}
\newblock
\shownote{Series Title: Lecture Notes in Computer Science.}


\bibitem[\protect\citeauthoryear{Guerraoui, Kuznetsov, Monti, Pavlovic,
  Seredinschi, and Vonlanthen}{Guerraoui et~al\mbox{.}}{2020}]%
        {guerraouiKMPSV2020}
\bibfield{author}{\bibinfo{person}{Rachid Guerraoui}, \bibinfo{person}{Petr
  Kuznetsov}, \bibinfo{person}{Matteo Monti}, \bibinfo{person}{Matej Pavlovic},
  \bibinfo{person}{Dragos-Adrian Seredinschi}, {and} \bibinfo{person}{Yann
  Vonlanthen}.} \bibinfo{year}{2020}\natexlab{}.
\newblock \showarticletitle{Scalable {Byzantine} {Reliable} {Broadcast}
  ({Extended} {Version})}.
\newblock \bibinfo{journal}{\emph{arXiv:1908.01738 [cs]}} (\bibinfo{date}{Feb.}
  \bibinfo{year}{2020}).
\newblock
\urldef\tempurl%
\url{https://doi.org/10.4230/LIPIcs.DISC.2019.22}
\showDOI{\tempurl}
\newblock
\shownote{arXiv: 1908.01738.}


\bibitem[\protect\citeauthoryear{Herlihy}{Herlihy}{2018}]%
        {Herlihy2018}
\bibfield{author}{\bibinfo{person}{Maurice Herlihy}.}
  \bibinfo{year}{2018}\natexlab{}.
\newblock \showarticletitle{Atomic cross-chain swaps}. In
  \bibinfo{booktitle}{\emph{Proceedings of the 2018 {ACM} symposium on
  principles of distributed computing}} \emph{(\bibinfo{series}{{PODC} '18})}.
  \bibinfo{publisher}{ACM}, \bibinfo{address}{New York, NY, USA},
  \bibinfo{pages}{245--254}.
\newblock
\showISBNx{978-1-4503-5795-1}
\urldef\tempurl%
\url{https://doi.org/10.1145/3212734.3212736}
\showDOI{\tempurl}
\newblock
\shownote{Number of pages: 10 Place: Egham, United Kingdom tex.acmid: 3212736.}


\bibitem[\protect\citeauthoryear{Herlihy, Liskov, and Shrira}{Herlihy
  et~al\mbox{.}}{2019}]%
        {HerlihyLS2021}
\bibfield{author}{\bibinfo{person}{Maurice Herlihy}, \bibinfo{person}{Barbara
  Liskov}, {and} \bibinfo{person}{Liuba Shrira}.}
  \bibinfo{year}{2019}\natexlab{}.
\newblock \showarticletitle{Cross-chain {Deals} and {Adversarial} {Commerce}}.
\newblock \bibinfo{journal}{\emph{Proceedings of the VLDB Endowment}}
  \bibinfo{volume}{13}, \bibinfo{number}{2} (\bibinfo{date}{Oct.}
  \bibinfo{year}{2019}), \bibinfo{pages}{100--113}.
\newblock
\showISSN{2150-8097}
\urldef\tempurl%
\url{https://doi.org/10.14778/3364324.3364326}
\showDOI{\tempurl}
\newblock
\shownote{arXiv: 1905.09743.}


\bibitem[\protect\citeauthoryear{Kalodner, Goldfeder, Chen, Weinberg, and
  Felten}{Kalodner et~al\mbox{.}}{2018}]%
        {Kalodner2018}
\bibfield{author}{\bibinfo{person}{Harry Kalodner}, \bibinfo{person}{Steven
  Goldfeder}, \bibinfo{person}{Xiaoqi Chen}, \bibinfo{person}{S.~Matthew
  Weinberg}, {and} \bibinfo{person}{Edward~W. Felten}.}
  \bibinfo{year}{2018}\natexlab{}.
\newblock \showarticletitle{Arbitrum: Scalable, private smart contracts}. In
  \bibinfo{booktitle}{\emph{27th USENIX Security Symposium (USENIX Security
  18)}}. \bibinfo{publisher}{USENIX Association}, \bibinfo{address}{Baltimore,
  MD}, \bibinfo{pages}{1353--1370}.
\newblock
\showISBNx{978-1-939133-04-5}
\urldef\tempurl%
\url{https://www.usenix.org/conference/usenixsecurity18/presentation/kalodner}
\showURL{%
\tempurl}


\bibitem[\protect\citeauthoryear{Kiayias, Russell, David, and
  Oliynykov}{Kiayias et~al\mbox{.}}{2017}]%
        {kiayias2017ouroboros}
\bibfield{author}{\bibinfo{person}{Aggelos Kiayias}, \bibinfo{person}{Alexander
  Russell}, \bibinfo{person}{Bernardo David}, {and} \bibinfo{person}{Roman
  Oliynykov}.} \bibinfo{year}{2017}\natexlab{}.
\newblock \showarticletitle{Ouroboros: {A} {Provably} {Secure}
  {Proof}-of-{Stake} {Blockchain} {Protocol}}.
\newblock In \bibinfo{booktitle}{\emph{Advances in {Cryptology} – {CRYPTO}
  2017}}, \bibfield{editor}{\bibinfo{person}{Jonathan Katz} {and}
  \bibinfo{person}{Hovav Shacham}} (Eds.). Vol.~\bibinfo{volume}{10401}.
  \bibinfo{publisher}{Springer International Publishing},
  \bibinfo{address}{Cham}, \bibinfo{pages}{357--388}.
\newblock
\showISBNx{978-3-319-63687-0 978-3-319-63688-7}
\urldef\tempurl%
\url{https://doi.org/10.1007/978-3-319-63688-7_12}
\showDOI{\tempurl}
\newblock
\shownote{Series Title: Lecture Notes in Computer Science.}


\bibitem[\protect\citeauthoryear{Kothapalli, Miller, and Borisov}{Kothapalli
  et~al\mbox{.}}{2017}]%
        {kothapalli2017smartcast}
\bibfield{author}{\bibinfo{person}{Abhiram Kothapalli}, \bibinfo{person}{Andrew
  Miller}, {and} \bibinfo{person}{Nikita Borisov}.}
  \bibinfo{year}{2017}\natexlab{}.
\newblock \showarticletitle{{SmartCast}: {An} {Incentive} {Compatible}
  {Consensus} {Protocol} {Using} {Smart} {Contracts}}.
\newblock In \bibinfo{booktitle}{\emph{Financial {Cryptography} and {Data}
  {Security}}}, \bibfield{editor}{\bibinfo{person}{Michael Brenner},
  \bibinfo{person}{Kurt Rohloff}, \bibinfo{person}{Joseph Bonneau},
  \bibinfo{person}{Andrew Miller}, \bibinfo{person}{Peter~Y.A. Ryan},
  \bibinfo{person}{Vanessa Teague}, \bibinfo{person}{Andrea Bracciali},
  \bibinfo{person}{Massimiliano Sala}, \bibinfo{person}{Federico Pintore},
  {and} \bibinfo{person}{Markus Jakobsson}} (Eds.).
  Vol.~\bibinfo{volume}{10323}. \bibinfo{publisher}{Springer International
  Publishing}, \bibinfo{address}{Cham}, \bibinfo{pages}{536--552}.
\newblock
\showISBNx{978-3-319-70277-3 978-3-319-70278-0}
\urldef\tempurl%
\url{https://doi.org/10.1007/978-3-319-70278-0_34}
\showDOI{\tempurl}
\newblock
\shownote{Series Title: Lecture Notes in Computer Science.}


\bibitem[\protect\citeauthoryear{Lamport}{Lamport}{1998}]%
        {Lamport1998}
\bibfield{author}{\bibinfo{person}{Leslie Lamport}.}
  \bibinfo{year}{1998}\natexlab{}.
\newblock \showarticletitle{The part-time parliament}.
\newblock \bibinfo{journal}{\emph{ACM Transactions on Computer Systems}}
  \bibinfo{volume}{16}, \bibinfo{number}{2} (\bibinfo{date}{May}
  \bibinfo{year}{1998}), \bibinfo{pages}{133--169}.
\newblock
\showISSN{0734-2071, 1557-7333}
\urldef\tempurl%
\url{https://doi.org/10.1145/279227.279229}
\showDOI{\tempurl}


\bibitem[\protect\citeauthoryear{Lev-Ari, Spiegelman, Keidar, and
  Malkhi}{Lev-Ari et~al\mbox{.}}{2019}]%
        {lev2019fairledger}
\bibfield{author}{\bibinfo{person}{Kfir Lev-Ari}, \bibinfo{person}{Alexander
  Spiegelman}, \bibinfo{person}{Idit Keidar}, {and} \bibinfo{person}{Dahlia
  Malkhi}.} \bibinfo{year}{2019}\natexlab{}.
\newblock \showarticletitle{{FairLedger}: {A} {Fair} {Blockchain} {Protocol}
  for {Financial} {Institutions}}.
\newblock \bibinfo{journal}{\emph{arXiv:1906.03819 [cs]}} (\bibinfo{date}{June}
  \bibinfo{year}{2019}).
\newblock
\urldef\tempurl%
\url{http://arxiv.org/abs/1906.03819}
\showURL{%
\tempurl}
\newblock
\shownote{arXiv: 1906.03819.}


\bibitem[\protect\citeauthoryear{McMenamin, Daza, and Pontecorvi}{McMenamin
  et~al\mbox{.}}{2021}]%
        {mcmenamin2021achieving}
\bibfield{author}{\bibinfo{person}{Conor McMenamin}, \bibinfo{person}{Vanesa
  Daza}, {and} \bibinfo{person}{Matteo Pontecorvi}.}
  \bibinfo{year}{2021}\natexlab{}.
\newblock \showarticletitle{Achieving {State} {Machine} {Replication} without
  {Honest} {Players}}.
\newblock \bibinfo{journal}{\emph{arXiv:2012.10146 [cs]}} (\bibinfo{date}{May}
  \bibinfo{year}{2021}).
\newblock
\urldef\tempurl%
\url{http://arxiv.org/abs/2012.10146}
\showURL{%
\tempurl}
\newblock
\shownote{arXiv: 2012.10146.}


\bibitem[\protect\citeauthoryear{Mendes, Tasson, and Herlihy}{Mendes
  et~al\mbox{.}}{2014}]%
        {MendesHT2012}
\bibfield{author}{\bibinfo{person}{Hammurabi Mendes},
  \bibinfo{person}{Christine Tasson}, {and} \bibinfo{person}{Maurice Herlihy}.}
  \bibinfo{year}{2014}\natexlab{}.
\newblock \showarticletitle{Distributed {Computability} in {Byzantine}
  {Asynchronous} {Systems}}.
\newblock \bibinfo{journal}{\emph{arXiv:1302.6224 [cs]}} (\bibinfo{date}{June}
  \bibinfo{year}{2014}).
\newblock
\urldef\tempurl%
\url{http://arxiv.org/abs/1302.6224}
\showURL{%
\tempurl}
\newblock
\shownote{arXiv: 1302.6224.}


\bibitem[\protect\citeauthoryear{Miller, Bentov, Kumaresan, and McCorry}{Miller
  et~al\mbox{.}}{2017}]%
        {MillerBKM17}
\bibfield{author}{\bibinfo{person}{Andrew Miller}, \bibinfo{person}{Iddo
  Bentov}, \bibinfo{person}{Ranjit Kumaresan}, {and} \bibinfo{person}{Patrick
  McCorry}.} \bibinfo{year}{2017}\natexlab{}.
\newblock \showarticletitle{Sprites: {Payment} channels that go faster than
  lightning}.
\newblock \bibinfo{journal}{\emph{CoRR}}  \bibinfo{volume}{abs/1702.05812}
  (\bibinfo{year}{2017}).
\newblock
\urldef\tempurl%
\url{http://arxiv.org/abs/1702.05812}
\showURL{%
\tempurl}
\newblock
\shownote{arXiv: 1702.05812 tex.bibsource: dblp computer science bibliography,
  http://dblp.org tex.biburl: http://dblp.org/rec/bib/journals/corr/MillerBKM17
  tex.timestamp: Wed, 07 Jun 2017 14:41:18 +0200.}


\bibitem[\protect\citeauthoryear{Miller, Xia, Croman, Shi, and Song}{Miller
  et~al\mbox{.}}{2016}]%
        {miller2016}
\bibfield{author}{\bibinfo{person}{Andrew Miller}, \bibinfo{person}{Yu Xia},
  \bibinfo{person}{Kyle Croman}, \bibinfo{person}{Elaine Shi}, {and}
  \bibinfo{person}{Dawn Song}.} \bibinfo{year}{2016}\natexlab{}.
\newblock \showarticletitle{The Honey Badger of BFT Protocols}. In
  \bibinfo{booktitle}{\emph{Proceedings of the 2016 ACM SIGSAC Conference on
  Computer and Communications Security}} (Vienna, Austria)
  \emph{(\bibinfo{series}{CCS '16})}. \bibinfo{publisher}{Association for
  Computing Machinery}, \bibinfo{address}{New York, NY, USA},
  \bibinfo{pages}{31–42}.
\newblock
\showISBNx{9781450341394}
\urldef\tempurl%
\url{https://doi.org/10.1145/2976749.2978399}
\showDOI{\tempurl}


\bibitem[\protect\citeauthoryear{Moniz, Neves, Correia, and Verissimo}{Moniz
  et~al\mbox{.}}{2011}]%
        {moniz2008ritas}
\bibfield{author}{\bibinfo{person}{Henrique Moniz},
  \bibinfo{person}{Nuno~Ferreria Neves}, \bibinfo{person}{Miguel Correia},
  {and} \bibinfo{person}{Paulo Verissimo}.} \bibinfo{year}{2011}\natexlab{}.
\newblock \showarticletitle{RITAS: Services for Randomized Intrusion
  Tolerance}.
\newblock \bibinfo{journal}{\emph{IEEE Transactions on Dependable and Secure
  Computing}} \bibinfo{volume}{8}, \bibinfo{number}{1} (\bibinfo{year}{2011}),
  \bibinfo{pages}{122--136}.
\newblock
\urldef\tempurl%
\url{https://doi.org/10.1109/TDSC.2008.76}
\showDOI{\tempurl}


\bibitem[\protect\citeauthoryear{Moscibroda, Schmid, and
  Wattenhofer}{Moscibroda et~al\mbox{.}}{2006}]%
        {moscibroda2006selfish}
\bibfield{author}{\bibinfo{person}{Thomas Moscibroda}, \bibinfo{person}{Stefan
  Schmid}, {and} \bibinfo{person}{Roger Wattenhofer}.}
  \bibinfo{year}{2006}\natexlab{}.
\newblock \showarticletitle{When selfish meets evil: byzantine players in a
  virus inoculation game}. In \bibinfo{booktitle}{\emph{Proceedings of the
  twenty-fifth annual {ACM} symposium on {Principles} of distributed computing
  - {PODC} '06}}. \bibinfo{publisher}{ACM Press}, \bibinfo{address}{Denver,
  Colorado, USA}, \bibinfo{pages}{35}.
\newblock
\showISBNx{978-1-59593-384-3}
\urldef\tempurl%
\url{https://doi.org/10.1145/1146381.1146391}
\showDOI{\tempurl}


\bibitem[\protect\citeauthoryear{Nakamoto}{Nakamoto}{2009}]%
        {bitcoin}
\bibfield{author}{\bibinfo{person}{Satoshi Nakamoto}.}
  \bibinfo{year}{2009}\natexlab{}.
\newblock \bibinfo{title}{Bitcoin: {A} peer-to-peer electronic cash system}.
\newblock
\newblock
\urldef\tempurl%
\url{http://www.bitcoin.org/bitcoin.pdf}
\showURL{%
\tempurl}


\bibitem[\protect\citeauthoryear{Nolan}{Nolan}{2016}]%
        {tiersnolan}
\bibfield{author}{\bibinfo{person}{T. Nolan}.} \bibinfo{year}{2016}\natexlab{}.
\newblock \bibinfo{title}{Atomic swaps using cut and choose}.
\newblock
\newblock
\urldef\tempurl%
\url{https://bitcointalk.org/index.php?topic=1364951}
\showURL{%
\tempurl}


\bibitem[\protect\citeauthoryear{Pass and Shi}{Pass and Shi}{2017}]%
        {pass2017fruitchains}
\bibfield{author}{\bibinfo{person}{Rafael Pass} {and} \bibinfo{person}{Elaine
  Shi}.} \bibinfo{year}{2017}\natexlab{}.
\newblock \showarticletitle{{FruitChains}: {A} {Fair} {Blockchain}}. In
  \bibinfo{booktitle}{\emph{Proceedings of the {ACM} {Symposium} on
  {Principles} of {Distributed} {Computing}}}. \bibinfo{publisher}{ACM},
  \bibinfo{address}{Washington DC USA}, \bibinfo{pages}{315--324}.
\newblock
\showISBNx{978-1-4503-4992-5}
\urldef\tempurl%
\url{https://doi.org/10.1145/3087801.3087809}
\showDOI{\tempurl}


\bibitem[\protect\citeauthoryear{Poon and Dryja}{Poon and Dryja}{2016}]%
        {lightning}
\bibfield{author}{\bibinfo{person}{J. Poon} {and} \bibinfo{person}{T. Dryja}.}
  \bibinfo{year}{2016}\natexlab{}.
\newblock \bibinfo{title}{The bitcoin lightning network: {Scalable} off-chain
  instant payments}.
\newblock
\newblock
\urldef\tempurl%
\url{https://lightning.network/lightning-network-paper.pdf}
\showURL{%
\tempurl}


\bibitem[\protect\citeauthoryear{Qasse, Abu~Talib, and Nasir}{Qasse
  et~al\mbox{.}}{2019}]%
        {qasse2019inter}
\bibfield{author}{\bibinfo{person}{Ilham~A Qasse}, \bibinfo{person}{Manar
  Abu~Talib}, {and} \bibinfo{person}{Qassim Nasir}.}
  \bibinfo{year}{2019}\natexlab{}.
\newblock \showarticletitle{Inter blockchain communication: A survey}. In
  \bibinfo{booktitle}{\emph{Proceedings of the ArabWIC 6th Annual International
  Conference Research Track}}. \bibinfo{pages}{1--6}.
\newblock
\urldef\tempurl%
\url{https://dl.acm.org/doi/pdf/10.1145/3333165.3333167}
\showURL{%
\tempurl}


\bibitem[\protect\citeauthoryear{Robinson}{Robinson}{2021}]%
        {robinson2021survey}
\bibfield{author}{\bibinfo{person}{Peter Robinson}.}
  \bibinfo{year}{2021}\natexlab{}.
\newblock \showarticletitle{Survey of crosschain communications protocols}.
\newblock \bibinfo{journal}{\emph{Computer Networks}}  \bibinfo{volume}{200}
  (\bibinfo{year}{2021}), \bibinfo{pages}{108488}.
\newblock
\urldef\tempurl%
\url{https://arxiv.org/pdf/2004.09494.pdf}
\showURL{%
\tempurl}


\bibitem[\protect\citeauthoryear{Rosu and Saleh}{Rosu and Saleh}{2019}]%
        {rocsu2021evolution}
\bibfield{author}{\bibinfo{person}{Ioanid Rosu} {and} \bibinfo{person}{Fahad
  Saleh}.} \bibinfo{year}{2019}\natexlab{}.
\newblock \showarticletitle{Evolution of {Shares} in a {Proof}-of-{Stake}
  {Cryptocurrency}}.
\newblock \bibinfo{journal}{\emph{SSRN Electronic Journal}}
  (\bibinfo{year}{2019}).
\newblock
\showISSN{1556-5068}
\urldef\tempurl%
\url{https://doi.org/10.2139/ssrn.3377136}
\showDOI{\tempurl}


\bibitem[\protect\citeauthoryear{Roughgarden}{Roughgarden}{2020}]%
        {roughgarden2020transaction}
\bibfield{author}{\bibinfo{person}{Tim Roughgarden}.}
  \bibinfo{year}{2020}\natexlab{}.
\newblock \showarticletitle{Transaction {Fee} {Mechanism} {Design} for the
  {Ethereum} {Blockchain}: {An} {Economic} {Analysis} of {EIP}-1559}.
\newblock \bibinfo{journal}{\emph{arXiv:2012.00854 [cs, econ]}}
  (\bibinfo{date}{Dec.} \bibinfo{year}{2020}).
\newblock
\urldef\tempurl%
\url{http://arxiv.org/abs/2012.00854}
\showURL{%
\tempurl}
\newblock
\shownote{arXiv: 2012.00854.}


\bibitem[\protect\citeauthoryear{Saleh}{Saleh}{2018}]%
        {saleh2021blockchain}
\bibfield{author}{\bibinfo{person}{Fahad Saleh}.}
  \bibinfo{year}{2018}\natexlab{}.
\newblock \showarticletitle{Blockchain {Without} {Waste}: {Proof}-of-{Stake}}.
\newblock \bibinfo{journal}{\emph{SSRN Electronic Journal}}
  (\bibinfo{year}{2018}).
\newblock
\showISSN{1556-5068}
\urldef\tempurl%
\url{https://doi.org/10.2139/ssrn.3183935}
\showDOI{\tempurl}


\bibitem[\protect\citeauthoryear{Sidhu}{Sidhu}{2020}]%
        {Sidhu2020}
\bibfield{author}{\bibinfo{person}{Jagdeep Sidhu}.}
  \bibinfo{year}{2020}\natexlab{}.
\newblock \bibinfo{title}{Blockchain {Bridges}, {Explained}}.
\newblock
\newblock
\urldef\tempurl%
\url{https://cointelegraph.com/explained/blockchain-bridges-explained}
\showURL{%
\tempurl}


\bibitem[\protect\citeauthoryear{Srikanth and Toueg}{Srikanth and
  Toueg}{1987}]%
        {SrikanthT1987}
\bibfield{author}{\bibinfo{person}{T.~K. Srikanth} {and} \bibinfo{person}{Sam
  Toueg}.} \bibinfo{year}{1987}\natexlab{}.
\newblock \showarticletitle{Simulating authenticated broadcasts to derive
  simple fault-tolerant algorithms}.
\newblock \bibinfo{journal}{\emph{Distributed Computing}} \bibinfo{volume}{2},
  \bibinfo{number}{2} (\bibinfo{date}{June} \bibinfo{year}{1987}),
  \bibinfo{pages}{80--94}.
\newblock
\showISSN{0178-2770, 1432-0452}
\urldef\tempurl%
\url{https://doi.org/10.1007/BF01667080}
\showDOI{\tempurl}


\bibitem[\protect\citeauthoryear{{Vitalik Buterin}}{{Vitalik Buterin}}{2021}]%
        {buterin2021}
\bibfield{author}{\bibinfo{person}{{Vitalik Buterin}}.}
  \bibinfo{year}{2021}\natexlab{}.
\newblock \bibinfo{title}{An {Incomplete} {Guide} to {Rollups}}.
\newblock
\newblock
\urldef\tempurl%
\url{https://vitalik.ca/general/2021/01/05/rollup.html}
\showURL{%
\tempurl}


\bibitem[\protect\citeauthoryear{Wattenhofer}{Wattenhofer}{2019}]%
        {sliwinski2020blockchains}
\bibfield{author}{\bibinfo{person}{Roger Wattenhofer}.}
  \bibinfo{year}{2019}\natexlab{}.
\newblock \bibinfo{title}{Blockchains {Cannot} {Rely} on {Honesty}}.
\newblock
\newblock
\urldef\tempurl%
\url{https://www.semanticscholar.org/paper/Blockchains-Cannot-Rely-on-Honesty-Wattenhofer/96b6a1c51a22dc658a3aaa84d333f51e5c8c4253}
\showURL{%
\tempurl}


\bibitem[\protect\citeauthoryear{Zakhary, Agrawal, and El~Abbadi}{Zakhary
  et~al\mbox{.}}{2019}]%
        {ZakharyAE2019}
\bibfield{author}{\bibinfo{person}{Victor Zakhary}, \bibinfo{person}{Divyakant
  Agrawal}, {and} \bibinfo{person}{Amr El~Abbadi}.}
  \bibinfo{year}{2019}\natexlab{}.
\newblock \showarticletitle{Atomic commitment across blockchains}.
\newblock \bibinfo{journal}{\emph{CoRR}}  \bibinfo{volume}{abs/1905.02847}
  (\bibinfo{year}{2019}).
\newblock
\urldef\tempurl%
\url{http://arxiv.org/abs/1905.02847}
\showURL{%
\tempurl}
\newblock
\shownote{arXiv: 1905.02847 tex.bibsource: dblp computer science bibliography,
  https://dblp.org tex.biburl:
  https://dblp.org/rec/bib/journals/corr/abs-1905-02847 tex.timestamp: Mon, 27
  May 2019 13:15:00 +0200.}


\end{thebibliography}

\end{document}